\documentclass[preprint,11pt]{aastex}
\tightenlines
\slugcomment{To appear in the Astrophysical Journal}

\newcommand    \aeff   {a_{\rm eff}}
\newcommand	\beq	{\begin{equation}}
\newcommand    \beqa   {\begin{eqnarray}}
\newcommand \cm	{\,{\rm cm}}
\newcommand	\eeq	{\end{equation}}
\newcommand    \eeqa   {\end{eqnarray}}
\newcommand	\eV	{\,{\rm eV}}
\newcommand	\g		{\,{\rm g}}
\newcommand	\gtsim	{\gtrsim}		 %apj version
\newcommand	\keV	{\,{\rm keV}}
\newcommand	\kpc	{\,{\rm kpc}}
\newcommand	\ltsim	{\lesssim}		 %apj version
\newcommand	\pc	{\,{\rm pc}}
\newcommand    \bahat  {{\hat{\bf a}}}
\newcommand    \bnhat  {{\hat{\bf n}}}
\newcommand    \bB     {{\bf B}}
\newcommand    \bJ     {{\bf J}}
\newcommand    \bJhat  {{\hat{\bf J}}}
\newcommand    \bmu    {{\bf\mu}}
\newcommand    \bxhat  {{\hat{\bf x}}}
\newcommand    \by     {{\bf y}}
\newcommand    \byhat  {{\hat{\bf y}}}
\newcommand    \bz     {{\bf z}}
\newcommand    \bzhat  {{\hat{\bf z}}}
\newcommand    \inc    {{\rm inc}}
\newcommand    \obs     {{\rm obs}}
\newcommand    \pmax           {p_{\rm max}}
\newcommand    \sca            {{\rm sca}}
\newcommand{\unitvec}[1]{\mathbf{\hat{#1}}}
\newcommand{\DADT}{\texttt{DADT}}
\newcommand{\MIEVzero}{\texttt{MIEV0}}

\begin{document}

\title{
%------------- enable for labelling preprint ---------------------------
%        \vspace*{-3.0em}
%        {\normalsize\rm v2.0 
%submitted to {\it The Astrophysical Journal}}\\ 
%}\\
        \vspace*{1.0em}
%-----------------------------------------------------------------------
	X-Ray Scattering by Nonspherical Grains. I. Oblate Spheroids
%	\\
%	{\small DRAFT: \today\ -- please do not circulate}
	}

\author{B.T. Draine and Khosrow Allaf-Akbari		%enter authors here
	}
\affil{Princeton University Observatory, Peyton Hall, Princeton,
NJ 08544; \\
{\tt draine@astro.princeton.edu, khosrow@astro.princeton.edu}}

\begin{abstract}
We calculate the scattering of X-rays by interstellar dust, for a dust model
that reproduces the 
observed wavelength-dependent extinction and polarization of starlight.
On interstellar sightlines that produce appreciable starlight polarization,
we predict that the dust-scattered
X-ray halo around point sources will have measurable azimuthal asymmetry
due to scattering by partially-aligned nonspherical grains.
We calculate the expected halo asymmetry.
X-ray halo asymmetry provides a new test of interstellar dust models.
\end{abstract}


\section{Introduction
	\label{sec:intro}}

Interstellar dust grains
scatter X-rays through small angles, as
was first pointed out by Overbeck (1965), Slysh (1969), and Hayakawa (1970). 
Because of this scattering, an image of an X-ray point source includes
a ``halo'' of X-rays that have been scattered
by dust grains near the line of sight.
First observed by Catura (1983) using the {\it Einstein} observatory,
scattered X-ray halos have since been measured by a number
of telescopes, including 
{\it Einstein} (e.g., Mauche \& Gorenstein 1986),
{\it ROSAT} (e.g., Predehl \& Schmitt 1995),
{\it Chandra} (e.g., Smith, Edgar, \& Shafer 2002),
and
{\it XMM-Newton} (e.g., Costantini, Freyberg, \& Predehl 2005).

The observed polarization of starlight (Hall 1949; Hiltner 1949) requires
that interstellar grains be both appreciably nonspherical and partially
aligned, with starlight propagating through the dusty interstellar medium
becoming linearly polarized as the result of preferential attenuation
of one of the linear polarization modes (``linear dichroism'').
In addition to producing polarization of starlight, aligned dust
grains produce polarized thermal emission at wavelengths from the far
infrared to the microwave.  Observers seeking to measure the polarization
of the cosmic microwave background must subtract 
this polarized ``galactic foreground'' from 
observations of the microwave sky.
Realistic models of nonspherical interstellar grains are therefore of
interest for many reasons.

There continue to be many uncertainties concerning the composition and
geometry of interstellar grains (for a recent review, see Draine 2003a),
and X-ray absorption and scattering can be used to test grain models.
We consider a specific grain model consisting of spherical
carbonaceous grains and oblate spheroidal silicate grains, with size
distributions and size-dependent degree of alignment adjusted to
reproduce
the observed wavelength dependence of both interstellar extinction
and interstellar polarization.

The scattering ``halo'' produced by a nonspherical grain will not be
azimuthally symmetric, and therefore the population of aligned interstellar
grains can be expected to produce asymmetric X-ray scattering halos.
Here we develop a method for calculating the differential scattering cross
section for X-rays incident on grains with arbitrary
geometry, and apply it to calculate X-ray scattering halos for a realistic
model of interstellar dust.

We find that the X-ray scattering halo produced by this model of aligned
intersellar grains has appreciable and observable asymmetry.
We propose statistics $R_\ell^{(I)}$ to measure the asymmetry of
observed halos.
For two models of partially-aligned interstellar grains, 
we calculate the expected values of $R_2^{(I)}$ and $R_4^{(I)}$, 
and discuss their observability.
The predicted values of $R_2^{(I)}$ 
should be measurable on sightlines where there
is appreciable polarization of starlight.
The ratio $R_4^{(I)}/R_2^{(I)}$ is sensitive to the shape of the
scattering grains. 

The paper is organized as follows:
In \S\ref{sec:x-ray scattering} we discuss anomalous diffraction theory,
as applied to interstellar grains, and our implementation of it.
In \S\ref{sec:spinning precessing grains} we discuss averaging over the
grain rotation expected for partially-aligned suprathermally-rotating grains.
In \S\ref{sec:azimuthal asymmetry} we introduce statistics $R_\ell^{(I)}$ 
to quantify azimuthal asymmetries in scattering halos.
We obtain a realistic grain model in \S\ref{sec:grain model}, and
in \S\ref{sec:anisotropic scattering by aligned dust} 
we use this model to predict
$R_2^{(I)}$ and $R_4^{(I)}$ 
for sightlines where the magnetic field direction is uniform and
perpendicular to the line-of-sight.
We discuss using halo asymmetry measurements as a test of
grain models in \S\ref{sec:discussion}.
Our results are summarized in \S\ref{sec:summary}.

A reader concerned only with the predicted observability of this
phenomenon may choose to proceed directly to \S\ref{sec:discussion}.

\section{\label{sec:x-ray scattering}
         X-Ray Scattering by Dust: Anomalous Diffraction Theory}

\subsection{Defining the Scattering Problem}

We consider an incident monochromatic electromagnetic plane wave 
with time dependence
$e^{-i\omega t}$ (not explicitly mentioned in the subsequent
formulation) and spatial dependence $\exp(i \mathbf{k\cdot r})$ 
 which is \textit{fully coherent} over the grain volume 
and \textit{totally non-coherent} over the length-scales of the
distances between them. 
The grain is assumed to
consist of material characterized by complex refractive index $m$. 
We neglect effects arising from the
crystalline or amorphous atomic structure of the grain 
(Bragg diffraction, for example, 
affects the scattering only at large angles, where
the scattering halo is impossible to observe).

An electromagnetic wave, in a homogeneous (and nonmagnetic) 
medium, would propagate as
\beq
\nabla^{2}{\mathbf{E}}(\mathbf{r}, \omega) +
k^{2}m^{2}{\mathbf{E}}(\mathbf{r}, \omega) = 0 ~~~.
\eeq
As usual, the properties of the vector field can be understood by studying the
behavior of each of its components, hence reducing our analysis to a
scalar theory of scattering, governed by a scalar wave equation 
$\nabla^{2}{U}(\mathbf{r}, \omega) +
k^{2}m^{2}{U}(\mathbf{r}, \omega) = 0$.
The difficulties arise from the coupling of the different vector components
by the boundary conditions at interfaces between different media.

The scattering problem hence reduces to finding how the  scatterer
responds to an incoming wave $U_\inc$, producing an outgoing wave
$U_\sca$, where $U = U_\inc + U_\sca$
should satisfy the wave equations above, as well as appropriate boundary
conditons.
For spheres, an exact series solution, first described by
Mie (1908) and Debye (1909) and
commonly referred
to as ``Mie theory'', can be employed provided the sphere is not too large
relative to the wavelength of the incident radiation.
However, nonspherical targets require other methods.

\subsection{Anomalous Diffraction Theory}

At X-ray energies materials have refractive indices 
very close to unity ($|m-1| \ll 1$)
and the dust grains responsible for most of the scattering
are usually much larger than the wavelength of the incoming radiation
($ka \gg 1 $). In this regime, the scattering and absorption of
X-rays can be calculated using an approximation first developed by
van de Hulst (1957) and known as
``anomalous diffraction theory'', hereafter
ADT.

ADT is a combination of ray-tracing optics
(applicable because $ka \gg 1$) 
and Huygens' principle of 
\textit{propagation of a scalar field},
applied to cases where the EM wave can
enter and propagate through the grain with essentially no reflection
or refraction ($|m-1| \ll 1$).

Consider an incident plane wave $U_\inc=U_0 e^{ikz}$, 
propagating in the $\bzhat$ direction.
Under the conditions above, the plane wave, once reaching
a plane \textbf{V} located just beyond the grain and normal
to the direction of the propagation of the incident wave, will have changed
by a fractional amount which we refer to as the 
\textit{shadow function}, $f(x,y)$:
\beq \label{eq:shadow function}
f(x, y) \equiv 1 - \exp[i\Phi(x, y)] ~~~,
\eeq
where the complex phase function $\Phi$ is
\beq
\Phi(x,y) \equiv k\int{[m(x, y, z)-1] dz} ~~~,
\eeq
where $m(x, y, z)$ is the refractive index at the
point $(x, y, z)$.

Once the shadow function is known, Huygens' principle allows the
amplitude of the scattered part of the wave, 
in the radiation zone, to be calculated as a Fourier
transform of the shadow function over the plane ${\bf V}$:
\begin{eqnarray}
U_\sca(r\unitvec{n}) &=& U_0\frac{\exp(ikr)}{kr}S(\unitvec{n})~~~,
\\
S( \bnhat )= S(k_x,k_y) &=& 
\frac{k^{2}}{2\pi} \int{ \exp[i (k_{x}x + k_{y}y)]f(x,y)dxdy} ~~~, 
\label{eq:S(nhat)}
\\
k_{x}&=& k(\unitvec{n} \cdot \unitvec{x}) ~~~,
\\ 
k_{y}&=& k(\unitvec{n} \cdot \unitvec{y}) ~~~.
\end{eqnarray} 

The scattering properties of the grain
can be obtained from $S(\bnhat)$, with the differential scattering
cross section given by:
\begin{eqnarray}
\frac{d\sigma_{sca}}{d\Omega}(\unitvec{n}) &=& 
\frac{|S(\unitvec{n})|^2}{k^{2}} ~~~,
\end{eqnarray}
The extinction cross section can be obtained from the optical theorem
[see, e.g., Bohren \& Huffman (1983)]
\begin{eqnarray}
\sigma_{\rm ext}&=&\frac{4\pi}{k^{2}}{\rm Re}[S(\unitvec{z})]
\\
%            &=& 2\int\left[1-\cos\Phi(x,y)\right]~dx~dy ~~~
%\\
            &=& 2\int\left(1-e^{-\Phi_2}\cos\Phi_1\right)~dx~dy ~~~
\end{eqnarray}
where $\Phi_1\equiv {\rm Re}(\Phi)$, $\Phi_2\equiv {\rm Im}(\Phi)$.
Radiation traversing the grain at $(x,y)$ is attenuated by a factor
$e^{-2\Phi_2}$, so that a fraction $(1-e^{-2\Phi_2})$ of the incident
power/area at $(x,y)$ is absorbed by the grain.
Thus the absorption cross section in the geometric optics
approximation assuming $|m-1|\ll1$ (reflection and refraction at interfaces
is small) is
\begin{eqnarray}
\sigma_{\rm abs} &=& 
\int \left( 1-e^{-2\Phi_2}\right)~dx~dy
~~~,
\end{eqnarray}
and the total scattering cross section 
$\sigma_{\rm sca}=\sigma_{\rm ext}-\sigma_{\rm abs}$ is
\begin{eqnarray}
\sigma_{\rm sca} &=&
\int \left[ 1-2\cos\Phi_1 e^{-\Phi_2} + 
e^{-2\Phi_2}\right]~dx~dy
\\
&=& \int |f|^2 ~dx~dy ~~~.
\end{eqnarray}
The dimensionless efficiency
factors for absorption, scattering, and extinction
are here defined to be 
$Q_{\rm abs}\equiv\sigma_{\rm abs}/\pi \aeff^2$,
$Q_\sca\equiv\sigma_\sca/\pi \aeff^2$,
$Q_{\rm ext}\equiv Q_{\rm abs}+Q_\sca=\sigma_{\rm ext}/\pi \aeff^2$,
where for a grain of solid volume $V$, the effective
radius $\aeff\equiv(3V/4\pi)^{1/3}$ is the radius of an equal-volume sphere.

\subsection{ADT for a Sphere}

For a sphere and scattering angle $\Theta$, 
the simple form of the shadow function allows $S(\Theta)$
to be written (van de Hulst 1957)
\beqa
S(\Theta)&=&(ka)^2\int_0^{\pi/2}
du \left(1-e^{-i\rho\sin u}\right)
J_0(ka\Theta\cos u)\sin u \cos u ~~~,
\\
\rho &\equiv& 2ka (m-1) ~~~,
\eeqa
where $J_0$ is the Bessel function of order 0.
The extinction, absorption, and scattering efficiency factors are
\beqa
Q_{\rm ext} &=& 2 + 
\frac
{4\left\{
\cos2\beta - e^{-\rho_2}
\left[
      \cos(\rho_1-2\beta)-|\rho|\sin(\rho_1-\beta)
\right]
 \right\}}
{|\rho|^2}
\\
Q_{\rm abs} &=& 1 + 
\frac{e^{-2\rho_2}}{\rho_2} + 
\frac{e^{-2\rho_2}-1}{2\rho_2^2}
\\
Q_{\rm sca}&=&Q_{\rm ext}-Q_{\rm abs}
\eeqa
where $\rho_1\equiv{\rm Re}(\rho)$, $\rho_2\equiv{\rm Im}(\rho)$,
and $\beta\equiv\arctan(\rho_2/\rho_1)$.

\subsection{The \DADT\ code}

\begin{figure}[h]
\plotone{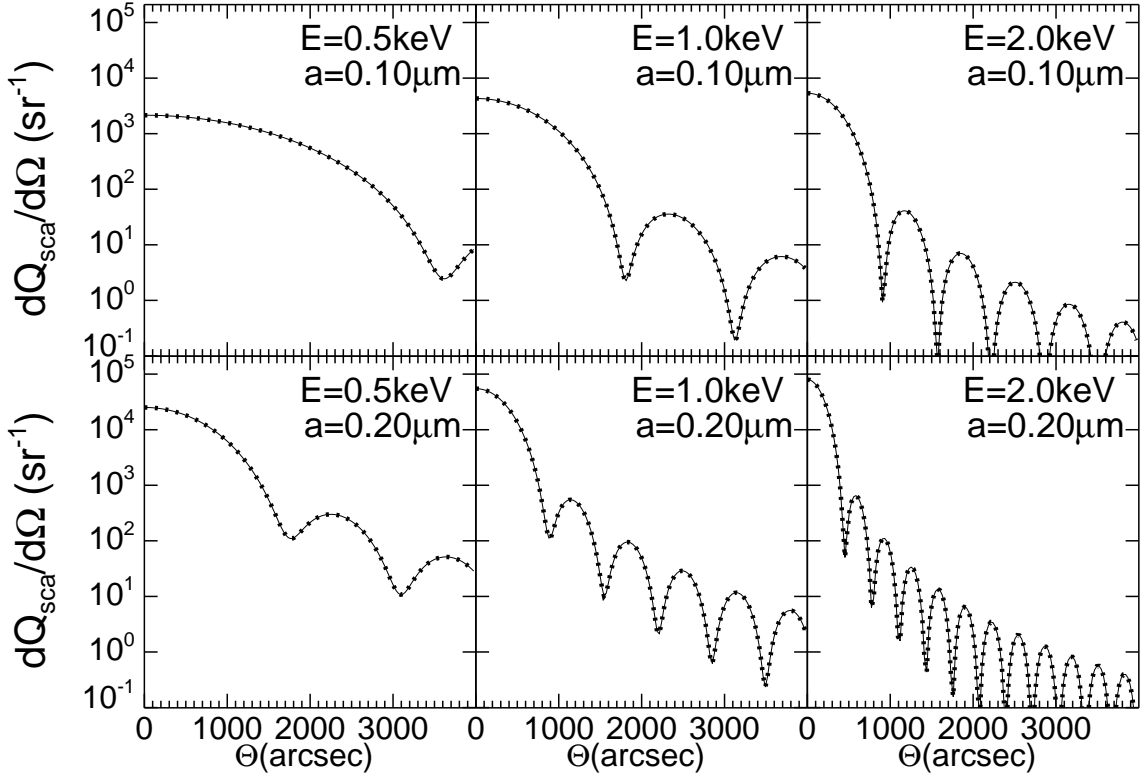}
 \caption{\label{fig:sphere}
   \footnotesize
   %f1.eps = fC1.eps:
   Differential scattering cross section for MgFeSiO$_4$
   silicate spheres with radii $a=0.1$ and $0.2\mu{\rm m}$ 
   at $E=0.5, 1$ and $2\keV$
   calculated with Mie theory (solid curve) and
   \DADT\ (dots).
   The two methods coincide to within the accuracy of the plot.
   The refractive index at $E=0.5, 1$ and $2\keV$ is taken to be
   $m=1-2.079\times10^{-3}+3.201\times10^{-3}i$,
   $1-7.152\times10^{-4}+1.887\times10^{-4}i$, and
   $1-1.920\times10^{-4}+2.807\times10^{-5}i$, respectively (Draine 2003b).
   }
  
 \end{figure}

For general shapes, eq.\ (\ref{eq:S(nhat)}) requires finding 
$f(x,y)$ numerically, 
followed by a two dimensional integration for each scattering
direction $\unitvec{n}$.
Because we will typically be interested in many scattering directions,
it is advantageous to employ fast Fourier transform (FFT) methods 
to find $S(\unitvec{n})$.

Our ADT-code, hereafter referred to as \texttt{Discrete-ADT}
code, \DADT\ for short, samples the shadow function on a 
$(x, y)$-grid of $2^{11} \times 2^{11}$ points, 
with $\sim 2^{7} \times 2^{7}$ 
points within the projected area of the grain.\footnote{
    The shadow function $f(x, y)$ is nonzero only over the 
    projection of the target onto the $(x, y)$-plane.
    However, any discrete Fourier transform
    requires that the range of integration be extended to distances 
    $r \gg a$ to evaluate $S(\unitvec{n})$ for angles within the first 
    minimum of the scattering halo.
    These angles play a crucial role in 
    our understanding of the halo properties, and we therefore
    extend our grid by a factor $2^4$ in each direction, providing
    an angular resolution of $\sim1/10$ the angle of the first
    minimum of $S(\bnhat)$.
    }
A 2-dimensional FFT then yields $S(k_x,k_y)$ over a rectangular 
$(k_x,k_y)$ grid of $2^{11}\times2^{11}$ points.
The single-precision
FFT code \texttt{GPFA}, developed by 
Temperton (1983, 1992) is employed for the 2-D FFT. 
The resulting $S(k_x,k_y)$
is then transformed  onto a $2^{11}\times 2^{11}$ \textsl{polar} lattice
(equal number of divisions in $\phi$ and $\Theta$) using a
two-dimensional cubic spline.

We have tested \DADT\ by comparing the scattering halo pattern
calculated 
using \DADT\ 
%for different silicate spheres for X-rays of energies between $0.5-2.0\keV$ 
to results calculated with the
Mie theory implementation \MIEVzero\ by Wiscombe (1980, 1996), after conversion
to double precision arithmetic.
Figure \ref{fig:sphere} shows the differential scattering cross section
calculated
for $a=0.1$ and $0.2\micron$ silicate spheres at $E=0.5, 1,$ and $2\keV$.
The results calculated with \DADT\ and those
calculated with \MIEVzero\ are indistinguishable.  
Because \MIEVzero\ and \DADT\ follow
entirely different approaches to the calculation, this confirms the accuracy
of both for the cases considered.
Note that absorption is strong in some of the cases shown:
${\rm Im}(m)ka=1.62$ for
$a=0.2\micron$ and $E=0.5\keV$.
ADT requires $|m-1|\ll 1$, but ${\rm Im}(m)ka$ need not be small.
The validity condition $|m-1\ltsim0.01$ is fulfilled for silicates for
$E>250\eV$.
The ray optics validity condition $ka \gtsim 10^2$ is satisfied for
$a > 0.08\micron (250\eV/E)$.

\begin{figure}[h]
\plotone{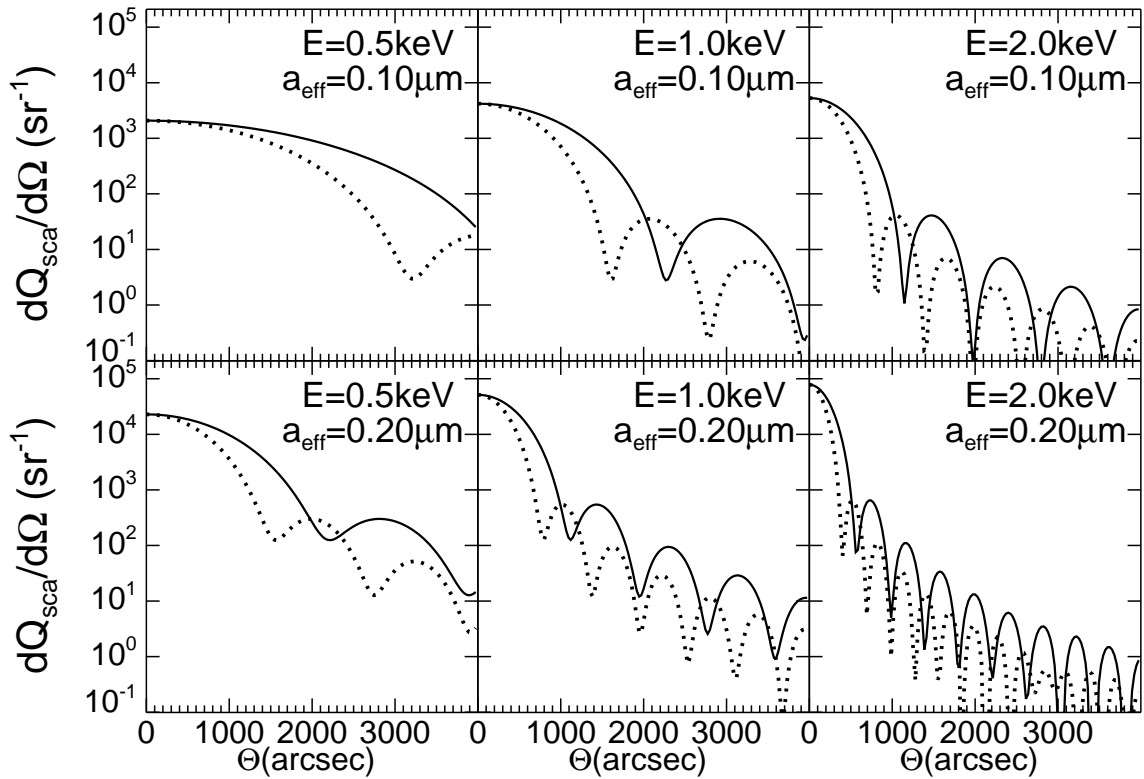}
\caption{
   \label{fig:dsigmadomega perp to los}
   \footnotesize
   Differential scattering cross section for 
   $b/a=\sqrt{2}$ oblate silicate spheroid with its symmetry-axis 
   perpendicular to the line of sight and in the direction $\phi=0$,
   for $\aeff = 0.1\micron$ and $0.2\micron$.   
   Results are shown for $\phi = 0^{\circ}$
   (solid line) and $\phi = 90^{\circ}$ (dotted line).
   Scattering is more extended in the direction parallel to
   the ``short axis'' of the shadow function.
   }
 \end{figure}

\begin{figure}[ht]
\plotone{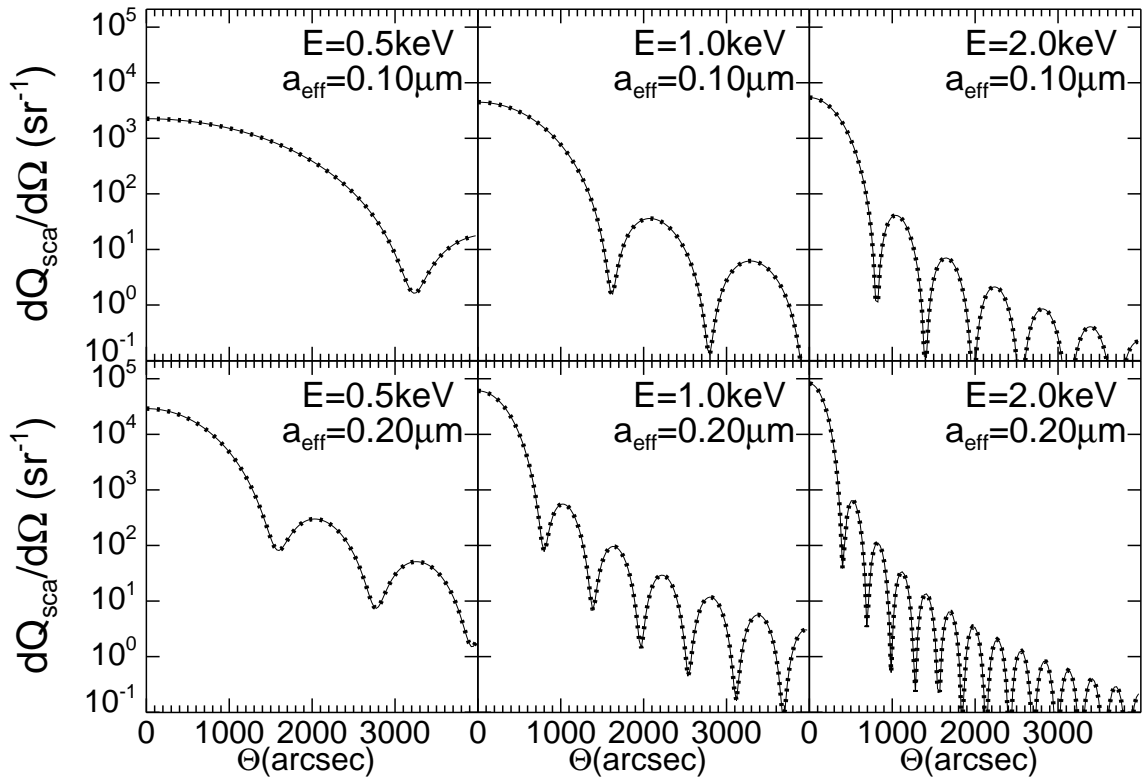}
\caption{
   \label{fig:dsigmadomega para to los}
   \footnotesize
   Same as Fig.\ \ref{fig:dsigmadomega perp to los},
   but for the grain symmetry axis parallel to the line of sight.
   Results are shown for
   $\phi = 0^{\circ}$ (solid line) and $\phi = 90^{\circ}$ (dotted line);
   the two curves coincide because of symmetry.
    }
 \end{figure}

\begin{figure}[h]
\plotone{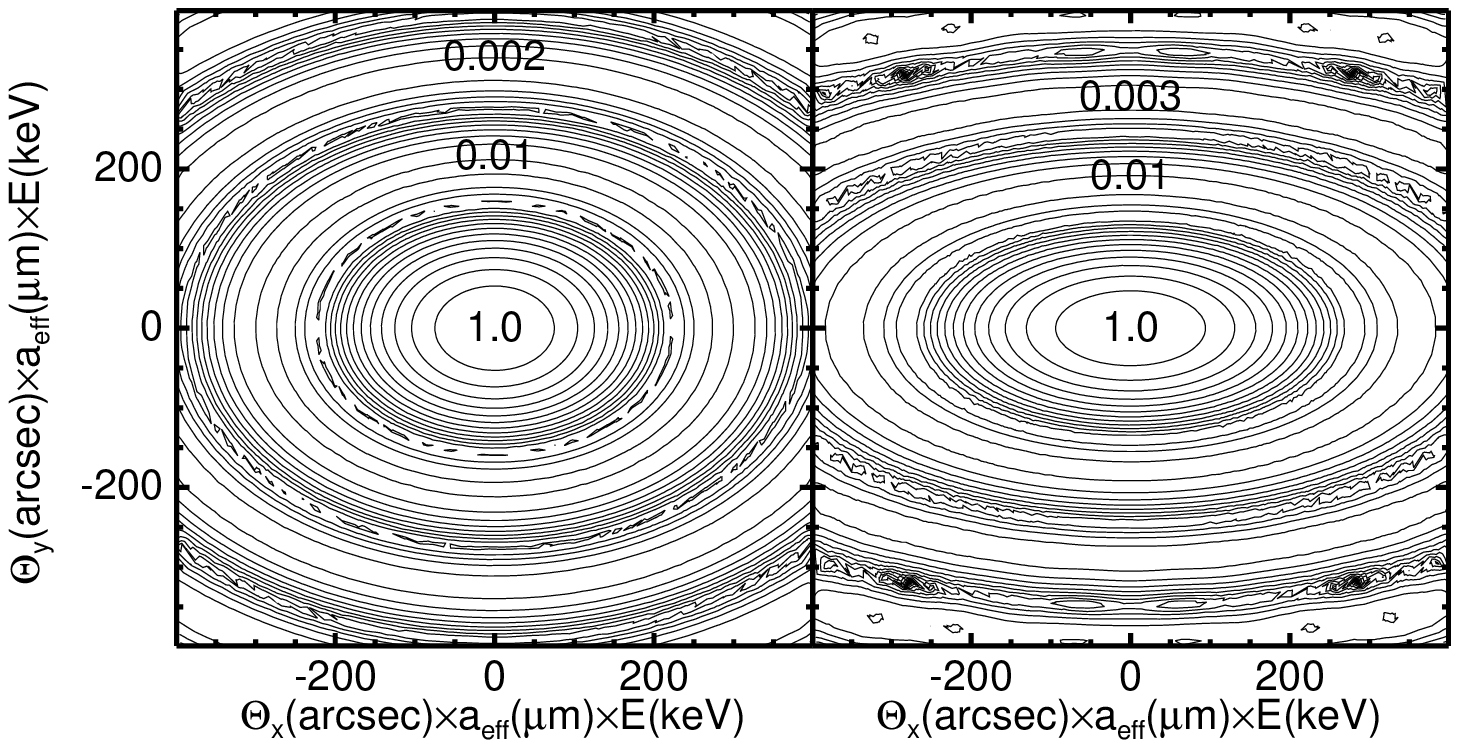}
\caption{\label{fig:halo images}
         \footnotesize
         Contours of constant 
	 $[d\sigma/d\Omega(\Theta,\phi)]/[d\sigma/d\Omega(0,0)]=10^{-n/5}$ 
	 for grains with
	 $b/a=\sqrt{2}$ (left panel) and
	 $b/a=2$ (right panel),
	 with short axis in the $\bxhat$ direction.
	 Contours at local maxima are labelled; regions where the
	 contours are closely-spaced are minima.
	 The contours should ideally be perfect ellipses; deviations result
	 from the discreteness of our grid and inaccuracies of interpolations
	 by the contour-plotting software.
	 Results were calculated for $\aeff=0.2\micron$ silicate grains
	 and $E=1\keV$, 
	 but the plots apply to other values of $\aeff$, $E$, and
	 composition provided
	 the ADT validity
	 criteria $|m-1|\ll 1$ and $ka\gg 1$ are satisfied
	 and, in addition, $|m-1|ka \ll 1$.
	 }
\end{figure}

\subsection{\label{sec:adt_results}
            Results}

Having verified that \DADT\ yields accurate results for X-ray scattering
by spheres, we now apply it to calculate X-ray scattering by oblate spheroids.
Figure \ref{fig:dsigmadomega perp to los} shows the differential scattering
cross section for axial ratio $b/a=\sqrt{2}$ 
oblate silicate spheroids with symmetry axis perpendicular
to the line-of-sight for two sizes ($\aeff=0.1$ and $0.2\micron$) and
three energies ($E=0.5, 1, 2\keV$).  The differential scattering cross 
section $d\sigma/d\Omega(\Theta,\phi)$ is shown as a function of $\Theta$
for $\phi=0$ (the direction of the short axis of the grain) 
and $\phi=90^\circ$.  It is apparent that the scattering is more extended
in the $\phi=0$ direction: it is easily shown that for this orientation of
an oblate spheroid, the halo extent in the $\phi=0$ direction is
larger than the extent in the $\phi=90^\circ$ direction by exactly the
factor $b/a$.

Figure \ref{fig:dsigmadomega perp to los} also shows that
as $\aeff$ or $E$ are increased, 
the angular extent of the halo shrinks, scaling as $1/(\aeff E)$.

Figure \ref{fig:dsigmadomega para to los}
shows the scattering for the same two grains as in Figure 2, but now
with the grains oriented with symmetry axis parallel to the line-of-sight.
For this orientation the scattering is azimuthally symmetric.
The angular extent (e.g., locations of minima and maxima) is similar
to the scattering in the $\phi=90^\circ$ direction in 
Fig.\ \ref{fig:dsigmadomega perp to los}, being proportional to $1/b$ in
both cases.

Figure \ref{fig:halo images} shows contours of constant $d\sigma/d\Omega$
for oblate spheroids, with symmetry axis (short axis) in the $\bxhat$
direction.  The contours are ellipses with axial ratios $b/a$.

\section{\label{sec:spinning precessing grains}
         Scattering by Spinning, Precessing Grains}

In this section we give a few general results concerning symmetries in
scattering by populations of spinning grains precessing around the
local magnetic field.
Rotation and precession are assumed to be
rapid enough to ensure phase averaging over rotation angle and precession
angle.

Consider radiation propagating in the $\bzhat$ direction.
Suppose that there is a magnetic field $\bB$ in the $\bxhat-\bzhat$ plane;
let $\Theta_{\bB\bz}$ be the angle between $\bB$ and $\bzhat$.
The scattering direction is defined by scattering angles $(\Theta,\phi)$,
where $\Theta$ is the deflection angle, and $\phi$ is an azimuthal angle
with $\phi=0$ corresponding to scattering in the $\bxhat-\bzhat$ plane.

Scattering by a sphere, or by a grain that is rotationally symmetric
about the line of sight, will be azimuthally symmetric.
This will also be the case for scattering by a population of 
arbitrarily-shaped grains if the distribution function for grain orientations
is
azimuthally symmetric about the line of sight.
 
Nonspherical grains that are not randomly oriented will produce scattering
that will not be azimuthally symmetric.
Grains will in general be spinning rapidly; let $\bJ$ be
the instantaneous direction of the grain angular momentum.
The torques acting on a spinning grain have been discussed elsewhere
(e.g., Draine \& Weingartner 1997).
The spinning grain will acquire a magnetic moment $\bmu\parallel-\bJ$
due to the Barnett effect.
The $\bmu\times\bB$ torque drives precession
of $\bJ$ around $\bB$.
The precession period is short, of order weeks, so that we may assume
a uniform distribution of $\bJ$ around the precession cone.
The distribution of directions $\bJhat$ will therefore be determined by 
$\Theta_{\bB\bz}$ and the angle $\Theta_{\bB\bJ}$ between $\bB$ and $\bJ$.

We will assume the aligned grains in the present study to
be spinning suprathermally, i.e., with rotational kinetic energy
$E_{\rm rot}\gg kT_{\rm gr}$, where $T_{\rm gr}$ is the grain temperature.
For a suprathermally-rotating grain with fixed angular momentum $\bJ$,
dissipation resulting from viscoelasticity, the Barnett effect, or
nuclear magnetism (Purcell 1979; Lazarian \& Draine 1999a,b)
will cause the grain to minimize its rotational kinetic energy,
and therefore to be oriented with 
$\bahat_1\parallel\bJ$ or $\bahat_1\parallel-\bJ$, 
where $\bahat_1$ is the principal axis of largest
moment of inertia.    
The states with
$\bahat_1\parallel\bJ$ and $\bahat_1\parallel-\bJ$ have the same energy,
and we will assume them to be equally occupied.
Therefore, for a suprathermally rotating grain, 
the distribution of $\bahat_1$ in
space will be fully determined by $\Theta_{\bB\bz}$ and the
alignment angle $\Theta_{\bB\bJ}$.

It is convenient to define a normalized scattering function
\beqa
\tilde{\sigma}(\Theta_s,\phi_s)&\equiv& \frac{1}{\sigma_\sca}
\langle \frac{d\sigma_\sca}{d\Omega}\rangle
\\
\sigma_\sca &\equiv& \int d\Omega ~\langle\frac{d\sigma_\sca}{d\Omega}\rangle
~~~,
\eeqa
where ($\Theta_s,\phi_s$) are scattering angles, 
$\langle d\sigma_\sca/d\Omega \rangle$
is the differential scattering cross section per H nucleon
summed over the different grain types and
averaged over the ensemble of grain orientations,
and $\sigma_\sca$ is the total scattering cross section per H nucleon.
With this definition, $\int \tilde{\sigma} d\Omega = 1$.
For suprathermally-rotating grains, the
normalized scattering function $\tilde{\sigma}(\Theta,\phi)$
is determined by the grain properties (composition, size, and geometry),
the angle $\Theta_{\bB\bz}$ and the
distribution function for the alignment angle $\Theta_{\bB\bJ}$ for
each grain type.

The function $\tilde{\sigma}$ will satisfy the symmetry\footnote{
   Eq. (\ref{eq:symmetry_1}) need not be satisfied for arbitrary grains, 
   but will be satisfied if every grain in the population has a mirror-image
   counterpart.}
\beq \label{eq:symmetry_1}
\tilde{\sigma}(\Theta,\phi)=\tilde{\sigma}(\Theta,-\phi)
\eeq
and can therefore be written
\beqa
\tilde{\sigma}(\Theta,\phi)
&=&
a_0(\Theta)+
2\sum_{\ell=1}^\infty a_\ell(\Theta)\cos(\ell\phi) ~~~,
\\ \label{eq:a_ell}
a_\ell(\Theta) &\equiv&
\frac{1}{2\pi}\int_0^{2\pi} d\phi ~ \tilde{\sigma}(\Theta,\phi)\cos(\ell\phi)
~~~.
\eeqa

If $\Theta_{\bB\bz}=0$ or $\pi/2$, the scattering will be symmetric
upon reflection through the $\by$ axis: 
\beqa
\label{eq:symmetry}
             \tilde{\sigma}(\Theta,\phi) &=&\tilde{\sigma}(\Theta,\pi-\phi)
~~~,
\\
a_\ell &=& 0 ~~~{\rm for~odd~}\ell .
\eeqa

For $0<\Theta_{\bB\bz}<\pi/2$, the symmetry (\ref{eq:symmetry}) is not
strictly required for scattering by dust.  However, when anomalous diffraction
theory applies, the symmetry condition (\ref{eq:symmetry}) 
will also apply,\footnote{
    This is a consequence of the scattering depending only on the
    shadow function (\ref{eq:shadow function}).
    For any $\bJ$, rotation of the grain around $\bJ$ and equal numbers
    of grains with $\bahat_1\parallel\bJ$ and $\bahat_1\parallel -\bJ$
    together ensure that the ensemble of shadow functions is symmetric under
    reflection $x \rightarrow -x$, in which case $\tilde{\sigma}(\Theta,\phi)=
    \tilde{\sigma}(\Theta,\pi-\phi)$.
    }
in which case
$\tilde{\sigma}(\Theta,\phi)$ can be written
\beq
\tilde{\sigma}(\Theta,\phi)
=
a_0(\Theta)+
2\sum_{\ell=1}^\infty a_{2\ell}(\Theta)\cos(2\ell\phi)
~~~.
\eeq
\begin{figure}[h]
\plotone{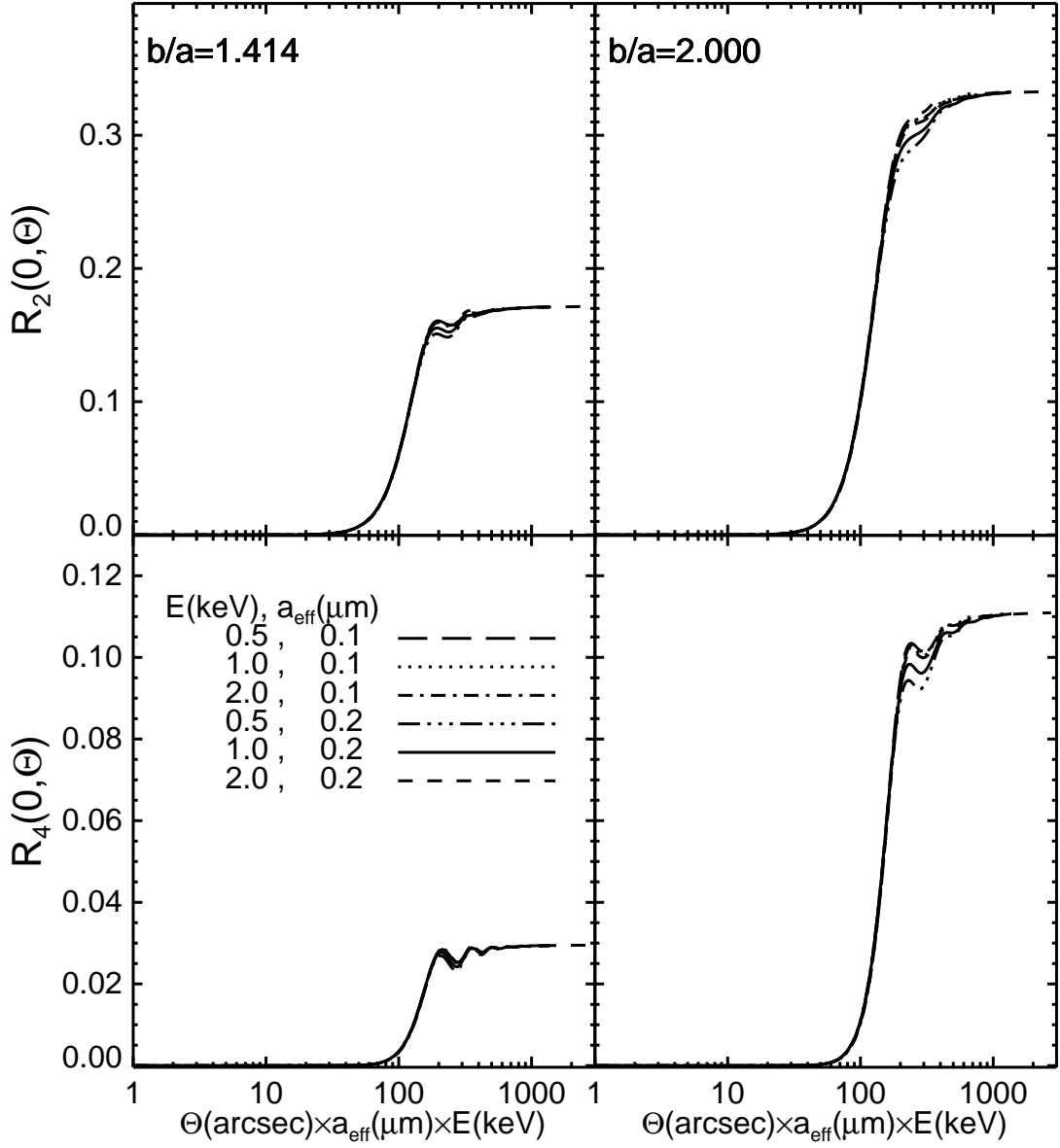}
\caption{\label{fig:R_2 and R_4 for E=0.5,1,2keV, a=0.1,0.2um}
         \footnotesize
	 $R_2^{(\sigma)}(0,\Theta)$ and $R_4^{(\sigma)}(0,\Theta)$ 
         vs.\ $\Theta\aeff E$
	 for $b/a=\sqrt{2}$ and $b/a=2$ silicate spheroids.
	 Results are shown for $\aeff=0.1$ and $0.2\micron$, and
	 for $E=0.5, 1, 2\keV$.
         }
\end{figure}

\section{\label{sec:azimuthal asymmetry}
         Azimuthal Asymmetry in X-Ray Scattering by a Nonspherical Grain}

\subsection{Asymmetry of Grain Scattering}

Consider some specific magnetic field direction $\Theta_{\bB\bz}$,
and an ensemble of grains
with some specified distribution function for the
alignment angle $\Theta_{\bB\bJ}$.
The function
\beqa
g^{(\sigma)}(\Theta_1,\Theta_2)&\equiv&
\int_{\Theta_1}^{\Theta_2}d\Theta\sin\Theta
\int_0^{2\pi}d\phi ~ \tilde{\sigma}(\Theta,\phi)
\\
&=& 2\pi \int_{\Theta_1}^{\Theta_2}d\Theta\sin\Theta a_0(\Theta)
\eeqa
gives the fraction of the scattering that is within the annulus
$[\Theta_1,\Theta_2]$.

For scattering angles in the annulus
$[\Theta_1,\Theta_2]$ 
the degree of azimuthal asymmetry of the differential
scattering cross section can be characterized by
functions
\beqa
R_\ell^{(\sigma)}(\Theta_1,\Theta_2)&\equiv&
\frac{\int_{\Theta_1}^{\Theta_2} d\Theta \sin\Theta \int_0^{2\pi}
                d\phi ~\cos (\ell\phi)
                ~\tilde{\sigma}(\Theta,\phi)}
{\int_{\Theta_1}^{\Theta_2} d\Theta\sin\Theta
 \int_0^{2\pi} d\phi ~ \tilde{\sigma}(\Theta,\phi)}
\\
&=& 
\frac{\int_{\Theta_1}^{\Theta_2} 
d\Theta \sin\Theta ~a_\ell(\Theta)}
{\int_{\Theta_1}^{\Theta_2} 
d\Theta \sin\Theta ~a_0(\Theta)}
~~~~~{\rm for}~\ell\geq 1
~~~.
\eeqa
The symmetry condition (\ref{eq:symmetry}) yields $R_\ell^{(\sigma)}=0$ for
odd $\ell$.

We will be interested primarily in the quadrupolar asymmetry,
characterized by the function $R_2^{(\sigma)}$.
Figure \ref{fig:R_2 and R_4 for E=0.5,1,2keV, a=0.1,0.2um}
shows $R_2^{(\sigma)}(0,\Theta)$ for
perfectly aligned oblate spheroids.
$R_2^{(\sigma)}(0,\Theta)$ rises rapidly from zero to a value
$\sim 0.17$ for $b/a=\sqrt{2}$ and $\sim0.33$ for $b/a=2$.
We also see that the octupole/quadrupole 
ratio $R_4^{(\sigma)}/R_2^{(\sigma)}$ is
sensitive to the grain shape: for 
$\Theta\aeff E \gtsim 3.5 ~{\rm arcmin}\,\micron\keV$, the ratio
$R_4^{(\sigma)}/R_2^{(\sigma)}$ increases from 0.17 to 0.33 
as $b/a$ is increased from $\sqrt{2}$ to 2.

\begin{figure}[t]
\begin{center}
\vspace*{-3em}
\includegraphics[angle=270,width=10cm]{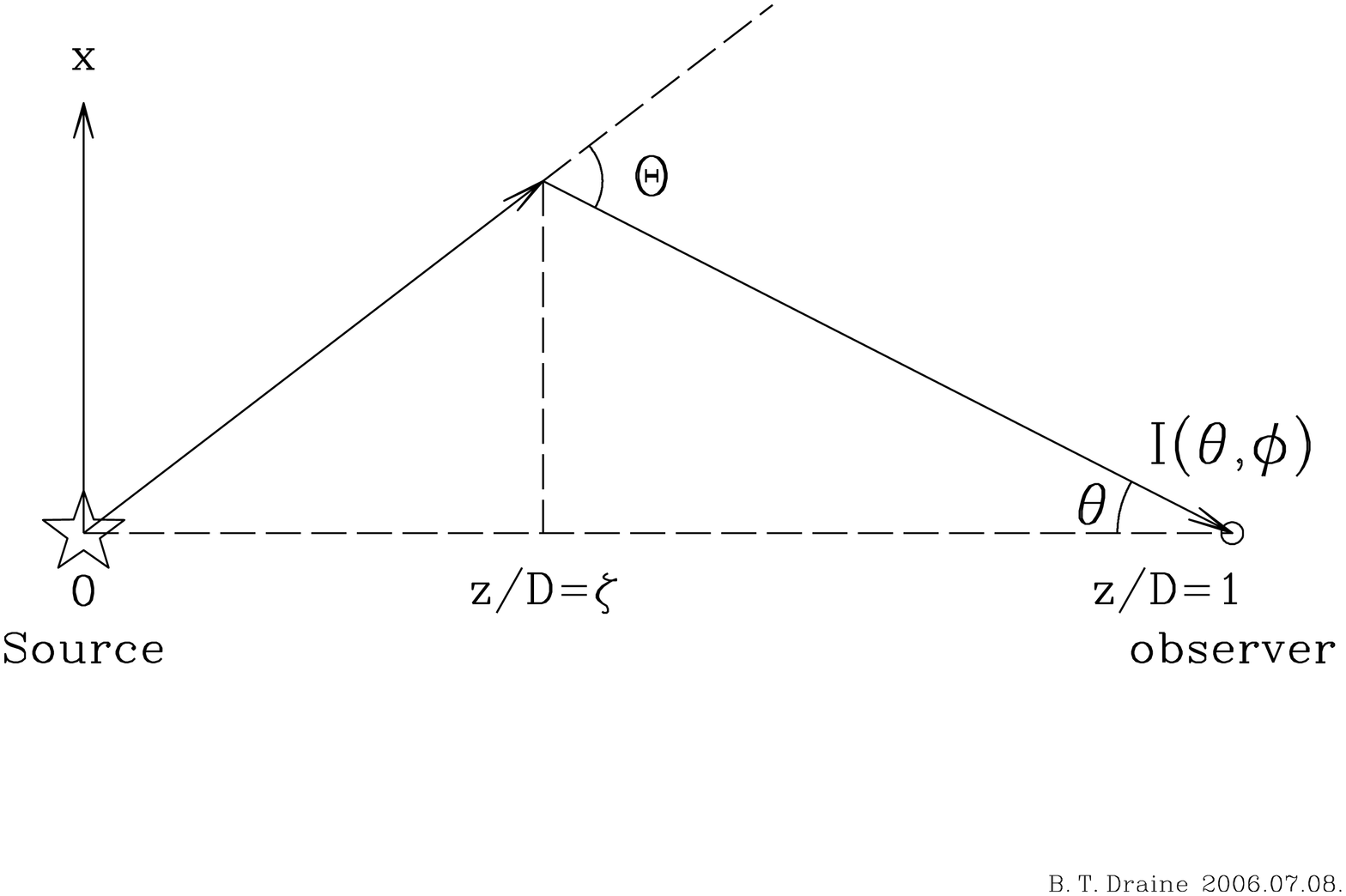}
\caption{\label{fig:geometry}\footnotesize
       Geometry for scattering of X-rays by a dust grain.  
       Angles are exaggerated;
       actual halo angles $\theta$ are $\ltsim 1^\circ$.
       }
\end{center}
\end{figure}

\subsection{Asymmetry of X-Ray Scattering Halo}

The geometry of X-ray scattering is illustrated in Figure \ref{fig:geometry};
note that X-ray scattering is significant only for small scattering angles
$\Theta$, and therefore only for small halo angles $\theta$.
If the dust grain distribution in space is $\rho(z)$ (assumed, for the moment,
to be independent of
$x$ and $y$ for small displacements from the line-of-sight), then, for
a steady isotropic source with luminosity per unit frequency $L_\nu$, the
specific intensity of singly-scattered photons is given by 
(see, e.g., Draine \& Tan 2000)
\beq
I_\nu(\theta,\phi)
\approx
\frac{L_\nu}{4\pi D^2} e^{-\tau_{\rm ext}} 
               \frac{1}{\cos\theta}
               \tau_\sca
  \int_0^1 d\zeta \frac{\tilde{\rho}(\zeta)~\tilde{\sigma}(\Theta_s,\phi)} 
                   {\zeta^2 + (1-\zeta)^2\tan^2\theta}
~~~,
\eeq
\beq
\tilde{\rho}(\zeta) \equiv \frac{\rho(z=\zeta D)}
                            {D^{-1}\int_0^D \rho(z)dz}
~~~,
\eeq
\beq
\Theta_s(\zeta,\theta)=\theta + 
\arctan\left[\frac{(1-\zeta)\tan\theta}{\zeta}\right]
~~~,
\eeq
\beq
\tau_\sca \equiv N_{\rm H} \sigma_\sca
~~~,
\eeq 
provided $\tau_\sca \ltsim 0.3$ so that multiple scattering can be neglected.
Let $I(\theta,\phi)$ be the observed intensity of scattered X-rays.
The function 
\beq
g^{(I)}(\theta_1,\theta_2)\equiv 
\frac{\int_{\theta_1}^{\theta_2}d\theta \sin\theta
      \int_0^{2\pi} d\phi ~I(\theta,\phi)}
     {\int_0^{\pi/2}d\theta \sin\theta
      \int_0^{2\pi} d\phi ~I(\theta,\phi)} ~~~,
\eeq
gives the fraction of the total scattered power that falls in the
annulus $[\theta_1,\theta_2]$.
%For optically-thin scattering, and neglecting scattering from dust very near
%the source, $g^{(I)}$ can be obtained from the 
%$g^{(\sigma)}$:
%\beq
%g^{(I)}(\theta_1,\theta_2) \approx 
%\int_{\theta_2}^1 d\zeta ~\tilde{\rho}(\zeta)~ 
%g^{(\sigma)}\left(\theta_1/\zeta,\theta_2/\zeta\right)
%~~~.
%\eeq
The azimuthal asymmetry of the scattered halo in the annulus
$[\theta_1,\theta_2]$ can be measured by
the function
\beq \label{eq:R_ell^(I)}
R_\ell^{(I)}(\theta_1,\theta_2) \equiv 
\frac{\int_{\theta_1}^{\theta_2} 
d\theta \sin\theta \int_0^{2\pi} d\phi ~
       \cos(\ell\phi)~I(\theta,\phi)}
     {\int_{\theta_1}^{\theta_2} 
d\theta \sin\theta \int_0^{2\pi} d\phi
       ~I(\theta,\phi)} ~~~.
\eeq
For optically-thin scattering by dust aligned by a uniform magnetic field
perpendicular to the line-of-sight, we can calculate $R_\ell^{(I)}$:
\beq \label{eq:R_ell^(I) in terms of a_ell}
R_\ell^{(I)}(\theta_1,\theta_2) 
     = \frac{\int_{\theta_1}^{\theta_2} d\theta \tan\theta
      \int_0^1 d\zeta ~a_\ell(\Theta_s)~\tilde{\rho}(\zeta)/
     [\zeta^2+(1-\zeta)^2\tan^2\theta]}
     {\int_{\theta_1}^{\theta_2} d\theta \tan\theta
      \int_0^1 d\zeta ~a_0(\Theta_s)~\tilde{\rho}(\zeta)
      /
     [\zeta^2+(1-\zeta)^2\tan^2\theta]} ~~~,
~~~
\eeq
where the $a_\ell$ are given by eq.\ (\ref{eq:a_ell}).
If we assume $\rho(\zeta)=0$ for 
$\zeta \ll 1$ (i.e., negligible scattering from
dust very near the source) we can assume $\tan\theta\ll 1$, 
$\theta\approx \zeta\Theta_s$, and
approximate (\ref{eq:R_ell^(I) in terms of a_ell}) by
\beq
R_\ell^{(I)}(\theta_1,\theta_2) \approx 
\frac{\int_0^1 d\zeta ~\tilde{\rho}(\zeta) 
      \int_{\theta_1/\zeta}^{\theta_2/\zeta}a_\ell(\theta)~\theta ~d\theta}
     {\int_0^1 d\zeta ~\tilde{\rho}(\zeta)
      \int_{\theta_1/\zeta}^{\theta_2/\zeta}a_0(\theta)~\theta~d\theta}
~~~.
\eeq

\section{\label{sec:grain model}
         Models for Aligned Interstellar Grains}
\subsection{Observed Polarization}

\begin{figure}[t]
\begin{center}
\includegraphics[width=12cm,angle=0]{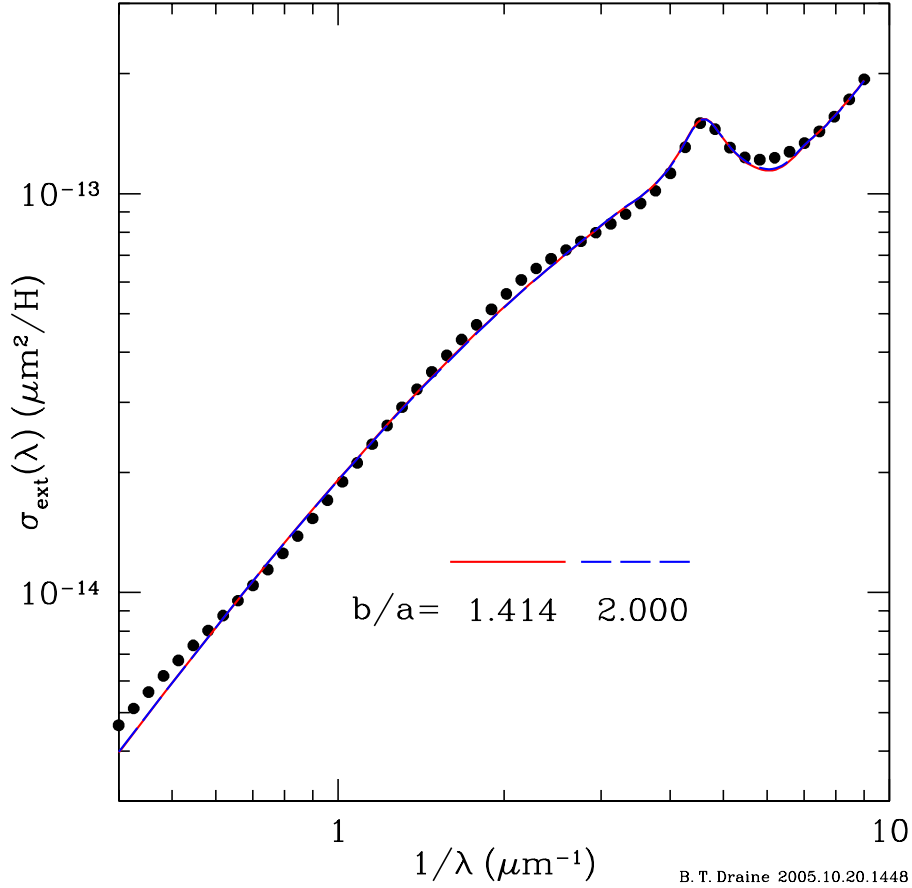}
%\plotone{f7.eps}
\caption{\label{fig:ext}
         \footnotesize
         Extinction for randomly-oriented grains with the
         size distribution shown in Figure \ref{fig:dnda}.
	 Dotted curve shows the observed extinction law that was
	 applied as a constraint on the model.
	 The extinction for the models with $b/a=\sqrt{2}$ (solid line)
	 and $2$ (broken line) nearly coincide and cannot be distinguished
	 in this plot.
        }
\end{center}
\end{figure}

Our objective is to calculate X-ray scattering from a realistic model
of partially aligned nonspherical dust grains, with the size distribution and
degree of alignment
constrained to reproduce both extinction and polarization as a function
of wavelength.

\begin{figure}[t]
\begin{center}
\includegraphics[width=12.0cm,angle=0]{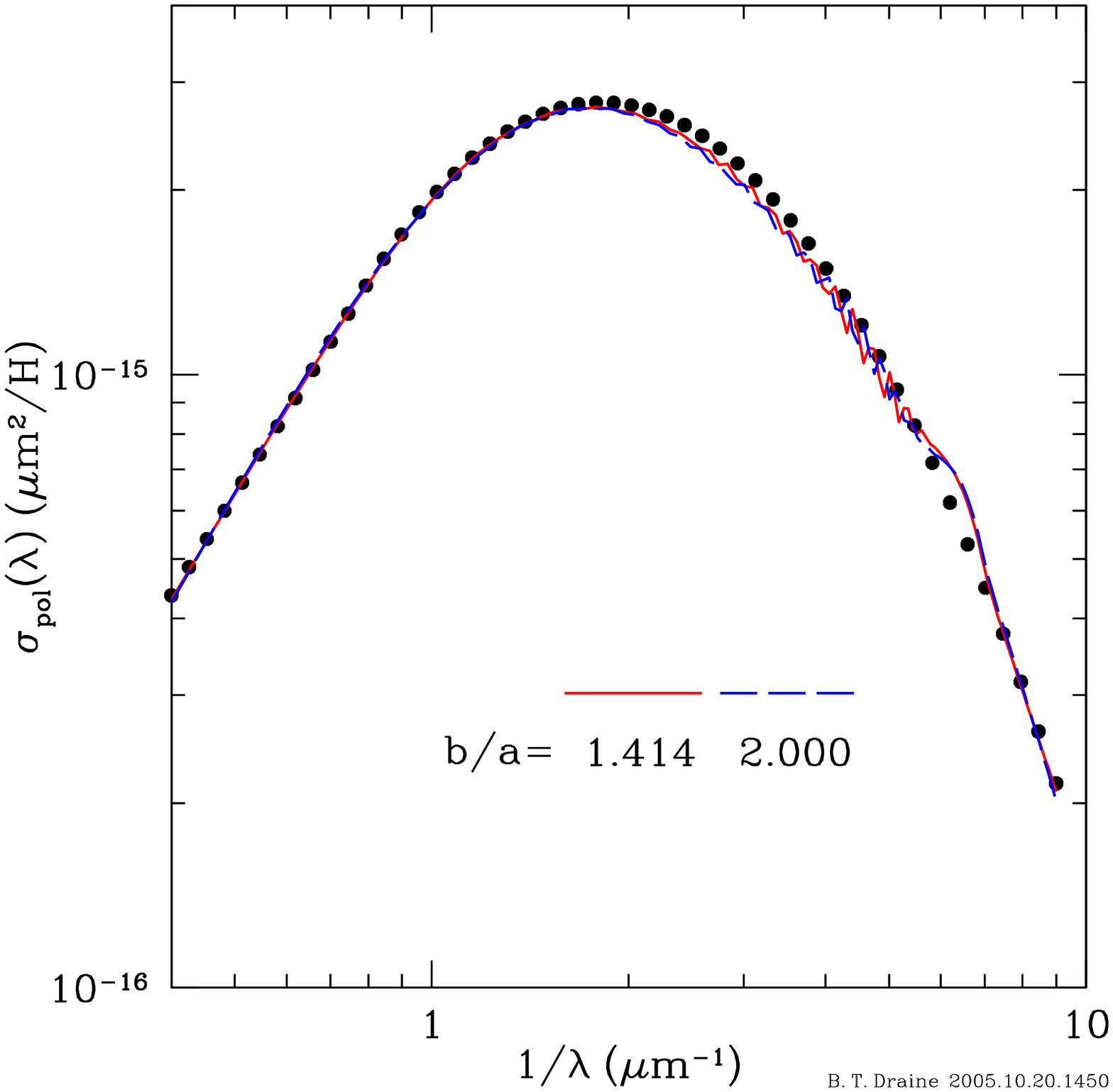}
\caption{\label{fig:pol}
         \footnotesize
	 %f8.eps :
         Polarization cross section $\sigma_{\rm pol}(\lambda)$ for
         a line of sight perpendicular to the local magnetic field,
	 for oblate silicate spheres with the size distribution of 
	 Fig.\ \ref{fig:dnda} and size-dependent alignment fraction
	 of Fig.\ \ref{fig:algn}.
	 Dotted curve is the observed polarization 
	 (eq.\ \ref{eq:Serkowski law},\ref{eq:IR polarization}) that
	 was applied as a constraint.  The models with
	 $b/a=\sqrt{2}$ (solid curve) and 2 (broken curve) nearly coincide
	 in this plot.
        }
\end{center}
\end{figure}

The observed polarization of starlight (Hall 1949; Hiltner 1949) demonstrates
that interstellar grains are both nonspherical and systematically aligned.
We take the polarization as a function of wavelength to be 
empirically described
by the ``Serkowski law'' (Serkowski 1973) for $\lambda < \lambda_x$, and
a power-law for $\lambda_x < \lambda < 5\micron$:
\beqa \label{eq:Serkowski law}
p(\lambda) &\approx& \pmax 
\exp\left\{ - K [\ln(\lambda/\lambda_{\rm max})]^2\right\}
~~~{\rm for}~ \lambda < \lambda_x 
~~~,
\\
\label{eq:IR polarization}
p(\lambda) &\approx& \pmax\exp\left(-\beta^2/4K\right)
(\lambda_x/\lambda)^\beta
~~~{\rm for}~ \lambda_x < \lambda < 5\micron
~~~,
\\
\lambda_x &=& \lambda_{\rm max}\exp(\beta/2K)
~~~.
\eeqa
The wavelength of peak polarization has a typical value
$\lambda_{\rm max}\approx 0.55\micron$, but varies
from one sightline to another.
The value of $K$ is correlated with $\lambda_{\rm max}$, with 
$K \approx 0.91(\lambda_{\rm max}/0.55\micron)+0.01$ (Whittet et al. 1992).
We take $\beta = 1.7$,
within the range observed by Martin et al.\ (1992).
For $\lambda_{\rm max}=0.55\micron$, 
we have $K=0.92$, and $\lambda_x=1.39\micron$.

Studies of many sightlines (Serkowski et al.\ 1975) find that
\beq
\pmax \ltsim 0.09 E(B-V)/{\rm mag}
\eeq
or
\beq
\pmax \ltsim 0.028 A(\lambda_{\rm max})/{\rm mag}
~~~.
\eeq
Sightlines with $\pmax/A(\lambda_{\rm max}) < .028/{\rm mag}$
are assumed to pass through regions where the magnetic field direction
is not transverse to the line-of-sight, or varies along the sightline,
or where the degree of grain alignment is for some reason lower than
average.

\subsection{Models With Partially-Aligned Dust Grains}

We model the dust as a mixture of carbonaceous particles (including PAHs)
and amorphous silicate particles.
Models of the infrared emission from interstellar dust
(Li \& Draine 2001) require a population of PAH particles containing
C$_{\rm PAH}$/H$_{\rm total}\approx 30-60$ppm.
The present models assume PAHs to be present with a
C$_{\rm PAH}$/H$_{\rm total}\approx 55$ppm, with optical properties
as described by Li \& Draine (2001).
In order to reproduce the observed wavelength-dependent extinction,
larger grains composed of both amorphous silicates and carbonaceous
materials are required (e.g., Mathis, Rumpl, \& Nordsieck 1977;
Draine \& Lee 1984; Weingartner \& Draine 2001; Zubko, Dwek, \& Arendt 2004).
The carbonaceous particles are here taken to be spherical or randomly-oriented
(and therefore not contributing to polarization), 
and the silicate particles are
assumed to be spheroids,
with diameter $2a$ along the symmetry axis, and $2b$ perpendicular to the
symmetry axis.
The size of a spheroid will be characterized by 
$\aeff\equiv(ab^2)^{1/3}$.

For the spherical carbonaceous grains, the IR to UV extinction 
was calculated using Mie theory, using the
dielectric tensor of graphite with the usual
1/3-2/3 approximation (Draine \& Malhotra 1993).
For the silicate grains, taken to be oblate spheroids,
we use the extended boundary condition method (EBCM) introduced by
Waterman (1971) and developed by
Mishchenko \& Travis (1994) and Wielaard et al (1997) -- see the
general review by Mishchenko, Travis \& Mackowski (1996).
Our computations make use of the code {\tt ampld.lp.f} 
(Mishchenko 2000).\footnote{
    {\tt ampld.lp.f} is available at
     http://www.giss.nasa.gov/$\sim$crmim/t\_matrix.html}
EBCM codes encounter computational difficulties when the target becomes
large compared to the wavelength.  
For the silicate oblate spheroids considered here,
the {\tt ampld.lp.f} code appeared to converge for $b/\lambda < 3.88$,
but sometimes failed for larger values of $b/\lambda$.
Thus for $b/\lambda < 3.88$ we used {\tt ampld.lp.f} to calculate
$Q_{\rm ext}$ for
different orientations, but took
\beq \label{eq:ebcm approx}
Q_{\rm ext}(b/\lambda,b/a,m) \approx Q_{\rm ext}(3.88,b/a,m)
~~~{\rm for}~b/\lambda > 3.88
~~~.
\eeq
For large values of $b/\lambda$, and for refractive indices appropriate
to the optical and ultraviolet, the extinction tends to be close to
twice the projected geometric cross section (with zero contribution to
polarization of starlight in this limit).
For the size distributions characteristic of interstellar
grains, grains with $b/\lambda > 3.88$ make only a minor contribution to
the total extinction, and a very small contribution to the polarization
of starlight, so the approximation (\ref{eq:ebcm approx}) does not
introduce significant error.

We consider models where all the silicate particles are oblate\footnote{
    There is some indication that oblate shapes provide a better match
    to polarization observations (Draine \& Lee 1984).
    Oblate spheroids also have the advantage of being 
    invariant under rotation about
    their principal axis of largest moment of inertia, thus
    eliminating the need for averaging over grain rotation.}
($b>a$) spheroids with a single
axial ratio $b/a$, independent of size.
The degree of alignment of grains of size $\aeff$ is given by an alignment
fraction $f(\aeff)$;
$0\leq f \leq 1$, where $f=0$ for random alignment, and $f=1$ for
grains where the short axis ($\bahat_1$)
is perfectly-aligned with the magnetic field
direction.
We assume the magnetic field to be perpendicular to the line-of-sight,
and for simplicity we approximate the distribution of partially-aligned
grains using ``picket fence alignment'': if the line-of-sight
is in the $\bzhat$ direction, and the magnetic field is in the
$\bxhat$ direction, we assume that a fraction $(1+2f)/3$ of the
oblate spheroids have $\bahat_1\parallel\bxhat$,
$(1-f)/3$ have $\bahat_1\parallel\byhat$, and 
$(1-f)/3$ have $\bahat_1\parallel\bzhat$.
This is not a physically realistic distribution -- 
as discussed in \S\ref{sec:spinning precessing grains},
one should properly integrate over
some distribution function $\phi(\Theta_{\bB\bJ})$ for the angle between
the grain angular momentum $\bJ$ and the magnetic field direction $\bB$.
However, given our lack of knowledge of the functional 
form of $\phi(\Theta_{\bB\bJ})$,
it is reasonable to assume picket-fence alignment, with a single number
$f(a)$ characterizing the alignment of grains of size $a$.

\begin{figure}[t]
\begin{center}
\includegraphics[width=12.0cm,angle=0]{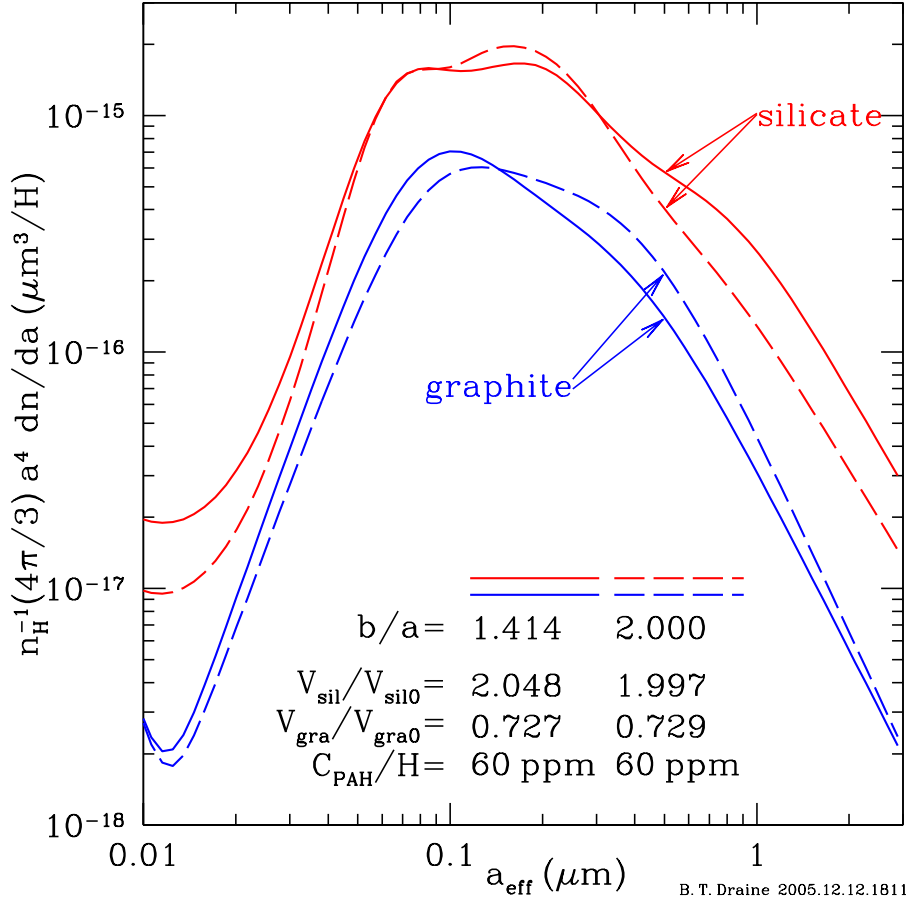}
\caption{\label{fig:dnda}
         \footnotesize
         %f9.eps :
         Size distributions for carbonaceous spheres and 
         amorphous silicate oblate spheroids.
	 Size distributions are constrained to be smooth, and
	 to reproduce the observed extinction and polarization.
	 The size distributions shown employ about twice as much
	 silicate material as would be permitted by current estimates of
	 solar abundances.
	 The size distributions are not shown for $\aeff < 0.01\micron$, 
	 because those 
	 grains contribute negligibly to X-ray scattering.
	 }
\end{center}
\end{figure}

\begin{figure}[t]
\begin{center}
\includegraphics[width=12.0cm,angle=0]{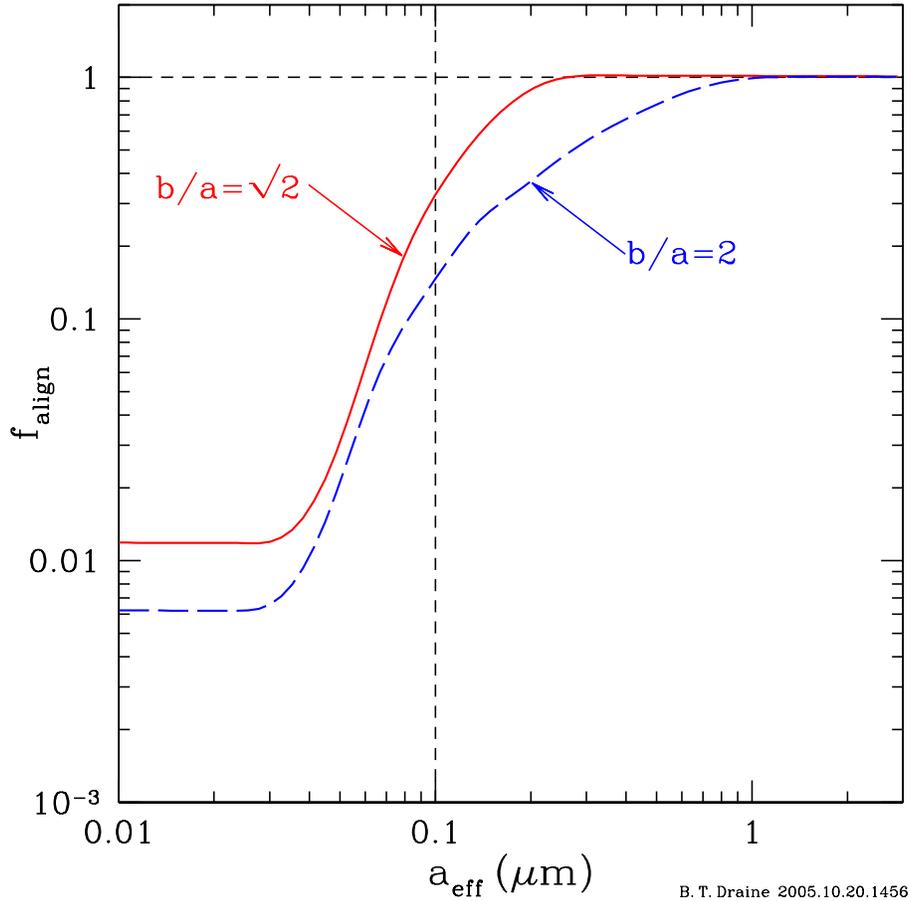}
\caption{\label{fig:algn}
         \footnotesize
	 %f10.eps :
         Fractional alignment for oblate silicate spheroids,
	 as a function of effective radius $\aeff$.
	 Large grains are perfectly aligned, with their principal
	 axis of largest moment of inertia parallel to the local
	 magnetic field direction.
	 Small grains are only minimally aligned.
        }
\end{center}
\end{figure}

We carry out a nonlinear least-squares fit for three continuous functions:
the size distributions
$(dn/da)_{\rm car}$ and $(dn/da)_{\rm sil}$ for carbonaceous grains
and silicate grains, and the alignment function $f(\aeff)$.
This fit is carried out using a number
of constraints, embodied in a penalty function $P=\sum_j (\Psi_j)^2$
that we seek to minimize.
One of the constraints is the requirement that
$p_{\rm max}/A(\lambda_{\rm max})=.028/$mag,
with $p(\lambda)/p_{\rm max}$ required to reproduce 
eq.\ (\ref{eq:Serkowski law},\ref{eq:IR polarization}).

In addition to a penalty for deviations from the observed extinction 
and polarization (at 100 wavelengths, logarithmically spaced from
$\lambda=2.5\micron$ to $\lambda=0.111\micron$),
the penalty function $P$ includes terms that are designed to favor
solutions for which the size distributions $dn_c/da$ and the
alignment function $f(a)$ are smooth functions of $a$.
In addition, the penalty function includes a term designed to
favor alignment functions $f(a)$ with $df/da \geq 0$,
as it is expected that small grains will be disaligned because of
the effects of ``thermal flipping'' (Lazarian \& Draine 1999a),
while starlight torques (Draine \& Weingartner 1997) will allow
large grains to overcome the effects of thermal flipping and achieve
suprathermal rotation, resulting in efficient
alignment of large grains
with the interstellar magnetic field in diffuse regions.

In addition, we include a penalty if the size distribution uses more than
the amount of silicate and carbon that would be present in solid form
in the ISM if the total interstellar abundances of C, Mg, Si, and Fe were
equal to current estimates of solar abundances.
Because we have no accepted theoretical expectation
for either the size distributions $dn/da$ or the alignment function $f(a)$,
the form of the penalty terms, and the weights they are given,
are necessarily somewhat arbitrary.  Our adopted penalty function
$P$ is described in the Appendix.

Kim \& Martin (1995) modeled the polarization of starlight
as a function of
wavelength using silicate spheroids.  
For oblate spheroids with axial ratio $b/a=\sqrt{2}$, they found
that the polarization can be explained if the grains with
$\aeff\gtsim 0.1\micron$ are nearly perfectly aligned, in which case 
a sightline perpendicular to $\bB$ will have polarization
$\pmax/A_V\approx 0.028/{\rm mag}$,
the maximum value of the ratio observed in the interstellar medium.
For larger values of $b/a$, only partial alignment of 
the $a\gtsim 0.1\micron$ silicate grains
is required to produce the observed polarization.

Following Kim \& Martin we take the silicate grains to be
oblate spheroids with axial
ratio $b/a=\sqrt{2}$, and we
find the size distributions $(dn/da)_{\rm car}$ and $(dn/da)_{\rm sil}$
and alignment function $f(\aeff)$ giving the best agreement with the
observational constraints.
We repeat this exercise for silicate grains with $b/a=2$.

Figure \ref{fig:ext} shows the extinction calculated for the two dust
mixtures, together with the observed extinction.
Figure \ref{fig:pol} shows the polarization as a function of wavelength
for the model, as well as the observed polarization.
Both models successfully reproduce the observed extinction and 
polarization.
However, in order to do so, 
both models require about twice as much silicate material
as would be allowed by current estimates for the solar abundances of
Mg, Si, and Fe.  Sofia \& Meyer (2001) recently discussed the
applicability of solar abundances to the interstellar medium,
and argue that abundances in young F and G stars provide a
better standard, with
Mg/H and Si/H about 12\% and 
23\% above the values currently favored for the Sun (although the
Mg and Si abundances in F and G stars have large uncertainties --
40\% and 33\%, respectively), but
this is still less than the amount of Mg and Si required
to reproduce the observed extinction and polarization for the
grain model considered here.
This abundance shortfall is generally encountered
by models that use ``compact'' grains to reproduce interstellar extinction
(e.g., Weingartner \& Draine 2001).
It may indicate that the abundances of interstellar Mg and Si are higher
than estimated from either the Sun or young F and G stars.
Alternatively, it may indicate that interstellar grains have other
geometries, e.g., ``composite'' grains with vacuum fractions of order
50\% or more (e.g., Zubko et al. 2004), which might permit the observed
extinction to be accounted for using less material in grains.

In Figure \ref{fig:dnda} we show the best-fit 
size distributions of silicate and
carbonaceous grains.
Figure \ref{fig:algn} shows the best-fit fractional alignment $f(\aeff)$.
As previously found by Kim \& Martin, silicate grains with $b/a=\sqrt{2}$
must be nearly perfectly aligned for $\aeff\gtsim 0.2\micron$ if the
observed polarization is to be reproduced.

\section{\label{sec:anisotropic scattering by aligned dust}
         X-Ray Scattering from 
	 Partially Aligned Interstellar Grains}

\subsection{Dust at a Single Distance}

We have calculated the X-ray scattering properties of graphite and silicate
dust grains with
the size distributions and fractional alignments
shown in Figs.\ \ref{fig:dnda} and 
\ref{fig:algn}, for 3 X-ray energies:
$E=0.5$, 1, and $2\keV$.
The total scattering cross section $\sigma_\sca$
is given in Table \ref{tab:sigma_sca} for the two partially-aligned
grain models, for two viewing directions: $\bz\parallel\bB$
and $\bz\perp\bB$.  The two models have similar values of
$\sigma_\sca$ -- the total strength of X-ray scattering does not appear
to discriminate between grain models with different degrees of grain
elongation, if the models are already constrained to reproduce the
observed optical extinction and polarization.

\begin{deluxetable}{l l c c c}
\tablewidth{0pt}
\tablecaption{\label{tab:sigma_sca}
                Total X-Ray Scattering Cross Section $\sigma_\sca$ 
		($10^{-24}\cm^2/{\rm H}$)\tablenotemark{a}}
%\begin{center}
\tablehead{\colhead{Model}&
           \colhead{Orientation} &
           \colhead{$E=0.5\keV$} & 
           \colhead{$E=1\keV$} & 
           \colhead{$E=2\keV$}
          }
\startdata
$b/a=1.414$ & $\bz\perp\bB$     & 163.  & 90.2   & 38.7 \\
  ``        & $\bz\parallel\bB$ & 153.  & 82.1   & 32.5 \\
  ``        & random            & 160.  & 87.5   & 36.7 \\
$b/a=2.000$ & $\bz\perp\bB$     & 172.  & 96.9   & 40.2 \\
``          & $\bz\parallel\bB$ & 145.  & 75.8   & 26.6 \\
``          & random            & 163.  & 89.9   & 35.7 \\
WD01 model\tablenotemark{b}  &random
                                & 173.  & 89.7   & 29.8\ \ \ \\
\enddata
\tablenotetext{a}{$\tau_\sca/A_V = (N_{\rm H}/A_V)\times\sigma_{\rm sca}$,
                  with $N_{\rm H}/A_V\approx 1.87\times10^{21}\cm^2/{\rm mag}$
		  (Bohlin et al.\ 1978).}
\tablenotetext{b}{$R_V=3.1$, C$_{\rm PAH}$/H=55ppm}
%\end{center}
\end{deluxetable}

\begin{figure}[t]
\begin{center}
\includegraphics[width=12.0cm,angle=0]{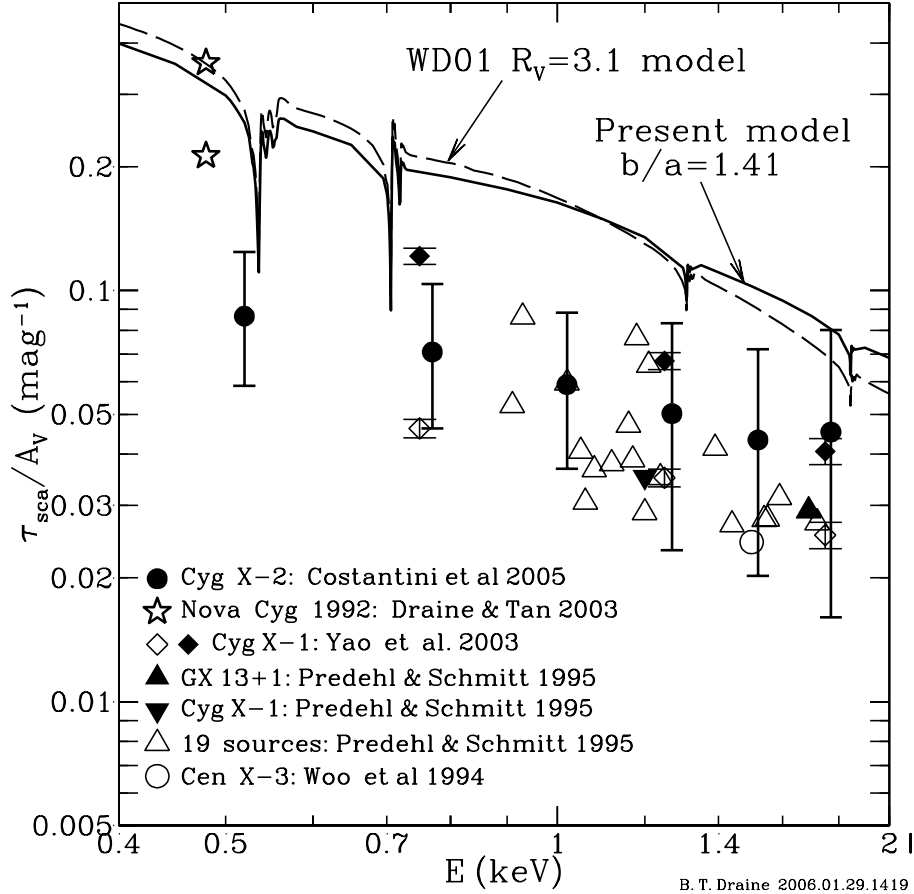}
\caption{\label{fig:tausca/AV vs E}
        \footnotesize
        %f11.eps = ftausca.eps :
	$\tau_{\rm sca}/A_V$ vs.\ energy $E$, where
	$\tau_{\rm sca}$ is the total optical depth for scattering,
	and $A_V$ is the visual extinction.
	Solid curve: scattering calculated for the present size distributions
	for graphite spheres and silicate spheroids (randomly-oriented) 
	with $b/a=1.414$ (results for $b/a=2$, not shown,
	are nearly identical).
	Broken curve: scattering calculated for graphite and silicate 
	spheres with 
	the WD01 size distribution.
	Also shown are observational results.  
	For the Cyg X-1 observations by Yao et al.\ (2003), the filled
	diamonds are corrected for scattering at $\theta>120\arcsec$
	(see Draine 2003b).
	For Nova Cyg 1992, results are shown for 
	$\tau_{\rm sca}=0.21$ (Draine \& Tan 2003) and two estimates of
	$E(B-V)$: 0.19~mag (Mathis et al.\ 1995) 
	and 0.32~mag (Vanlandingham et al.\ 2005).
	}
\end{center}
\end{figure}

Also shown in Table \ref{tab:sigma_sca} are X-ray scattering cross sections
calculated for a graphite+silicate grain model for spherical grains
with the size
distribution adopted by Weingartner \& Draine (2001; hereafter WD01).  
The WD01 size distribution is based on using graphite and silicate
spheres plus PAH molecules to reproduce the
observed 
interstellar extinction from the infrared to the ultraviolet; 
the resulting size distribution differs in detail 
from what is found here for spheroids contrainted to also reproduce
the polarization of starlight, but 
the total X-ray scattering
cross sections $\sigma_\sca$ are very similar to those obtained
here for spheroidal grains.

The ratio of X-ray scattering to visual extinction, $\tau_{\rm sca}/A_V$,
is plotted in Fig.\ \ref{fig:tausca/AV vs E} for both the present models
and the WD01 model, together with observational
determinations.
Aside from the apparent agreement between model and observation for
the soft X-rays from Nova Cyg 1992 (Draine \& Tan 2003), there appears
to be a general tendency for the observationally-determined X-ray
$\tau_{\rm sca}/A_V$ to be smaller than predicted by grain models based
on mixtures of solid carbonaceous grains and solid silicate grains.
However, 
measuring the X-ray scattering by dust is difficult at small angles, 
where it is mixed with
the point spread function, and at large angles, where background
corrections are uncertain.
At this time we consider the current
grain model to be viable, despite the fact that most of the observational
data in Fig.\ \ref{fig:tausca/AV vs E} appears to fall significantly
below the prediction.
Definitive measurement of the total X-ray
scattering cross section per unit $A_V$ would be of great value.

The total scattering cross sections given in Table \ref{tab:sigma_sca}
appears to
decrease by $\sim6-35\%$, depending on energy and $b/a$, 
when the viewing angle is changed from
$\bz\perp\bB$ (maximum optical polarization) to
$\bz\parallel\bB$ (zero optical polarization).
This small effect would probably be difficult to confirm observationally.

\begin{figure}[t]
\begin{center}
\includegraphics[width=18.0cm,angle=0]{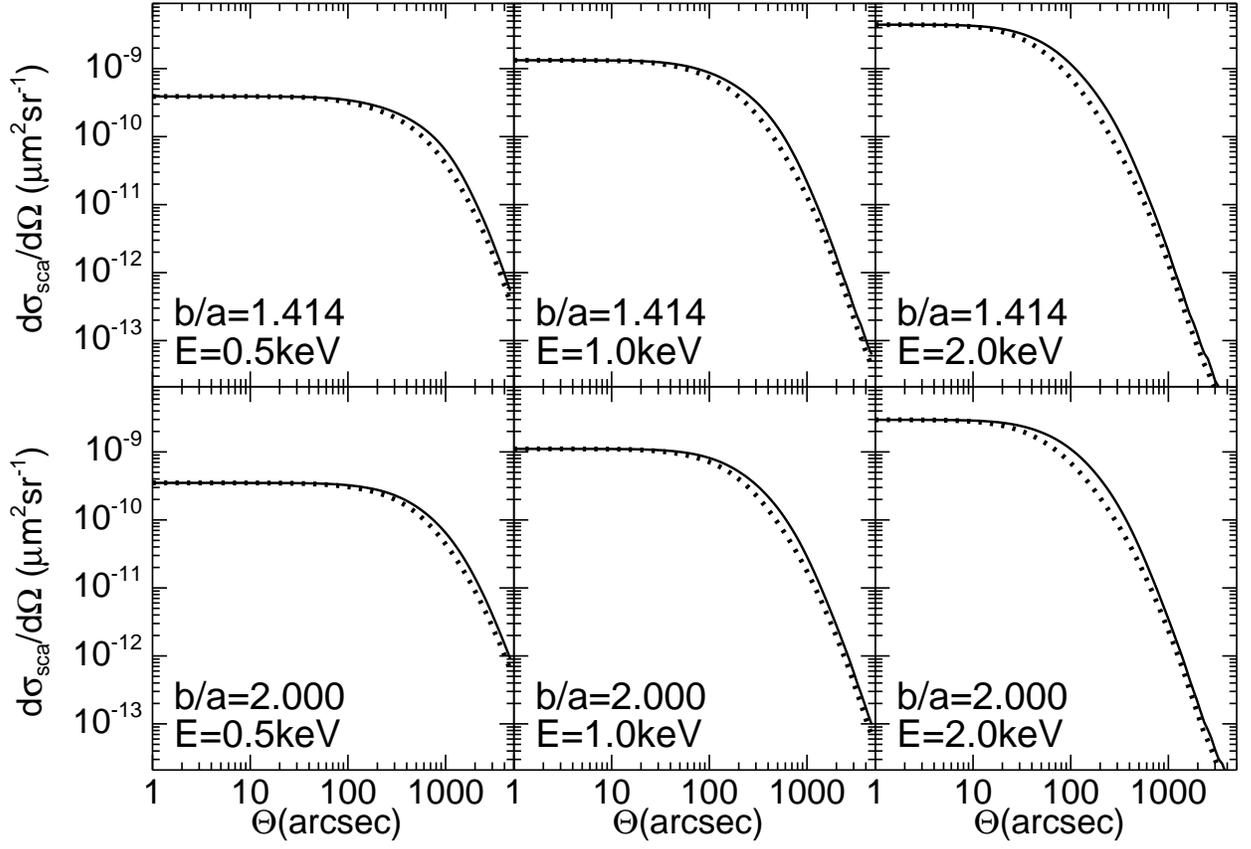}
\caption{\label{fig:dsigma/dOmega vs. theta for size dists.}
        \footnotesize
        Differential scattering cross sections per H nucleon
	vs.\ scattering angle $\Theta$ for
	the grain mixtures of Fig.\ \ref{fig:dnda}
	at $\phi=0$ (solid line) and $\phi=\pi/2$ (dotted line).
	Results are shown for
	$E=0.5,1,2\keV$, and for silicate grain axial ratio
	$b/a=\sqrt{2}$ and $b/a=2$.
	For optically-thin scattering by dust at a single distance,
	the halo intensity 
	$I(\theta,\phi)\propto d\sigma/d\Omega(\Theta,\phi)$,
	where $\Theta=\theta/\zeta$ for $\zeta\equiv z/D$, where
	$z$ is the distance from the source to the dust and 
	$D$ is the distance from the source to the observer.
	}
\end{center}
\end{figure}

\begin{figure}[ht]
\begin{center}
\includegraphics[width=16.0cm,angle=0]{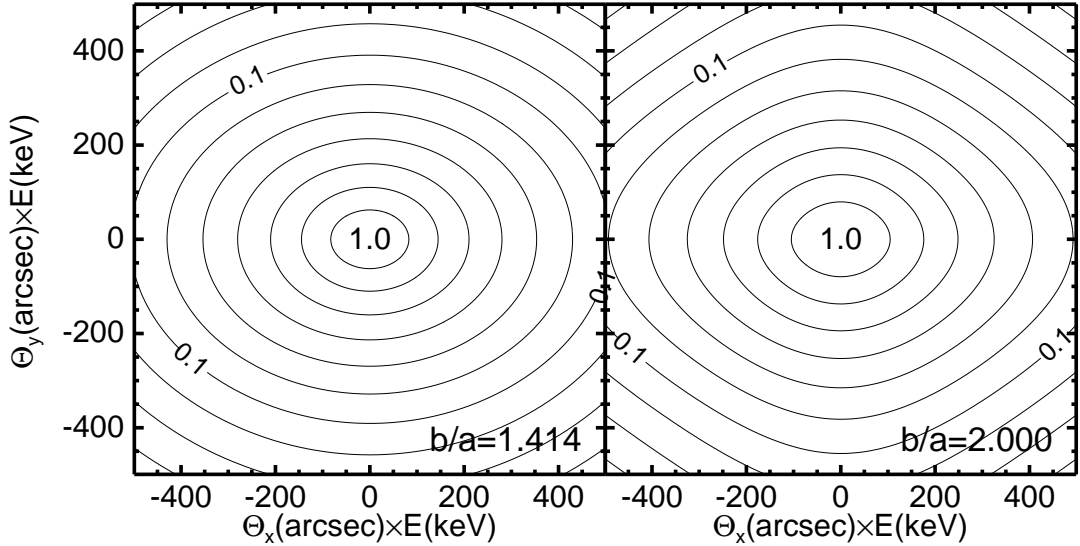}
\caption{\label{fig:image of sigmatilde}
        \footnotesize
        %f13.eps = fG13.eps:
        Contours of constant $\tilde{\sigma}(\Theta,\phi)/\tilde{\sigma}(0,0)
	= 10^{-n/7}$ 
	for $E=1\keV$
	for grain models including partially-aligned silicate grains
	with $b/a=\sqrt{2}$ and 2, for
	magnetic field $\bB \parallel \bxhat$.
        }
\end{center}
\end{figure}

\begin{figure}[ht]
\begin{center}
\includegraphics[width=10.0cm,angle=0]{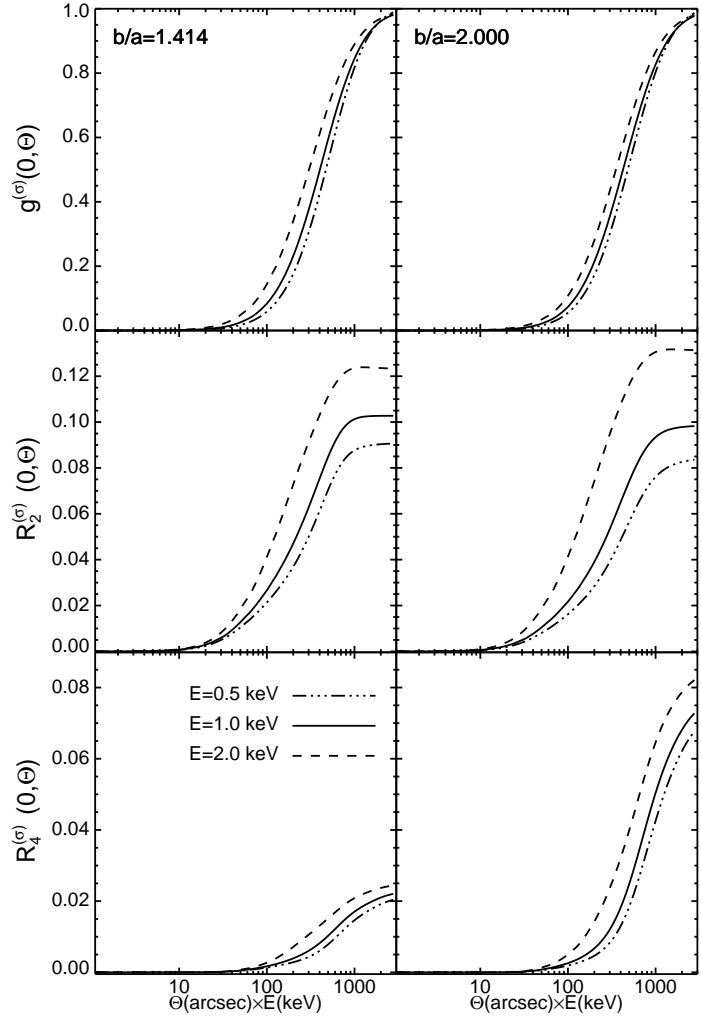}
\caption{\label{fig:R_2,R_4: 0.5,1,2keV; mix; sheet; b/a=1.4,2}
        \footnotesize
        $R_2^{(\sigma)}(0,\Theta)$ and $R_4^{(\sigma)}(0,\Theta)$ 
	vs.\ $\Theta E$ 
        for dust mixtures with realistic size distributions and degree of
	alignment.
	For
	grains in a sheet at a distance $(1-\zeta)D$ from the observer, where
	$D$ is the distance to the
        source, 
	$R_n^{(I)}(0,\theta)=R_n^{(\sigma)}(0,\theta=\zeta\Theta)$, 
	where $\theta$ is the
	observed halo angle.
	Results for $E=0.5,1,2\keV$
	are shown for grain models where the silicate grains are oblate
	spheroids with axial ratios
	$b/a=\sqrt{2}$ and 2.
	}
\end{center}
\end{figure}

%------------------- uniformly-distributed grains
\begin{figure}[ht]
\begin{center}
\includegraphics[width=16.0cm,angle=0]{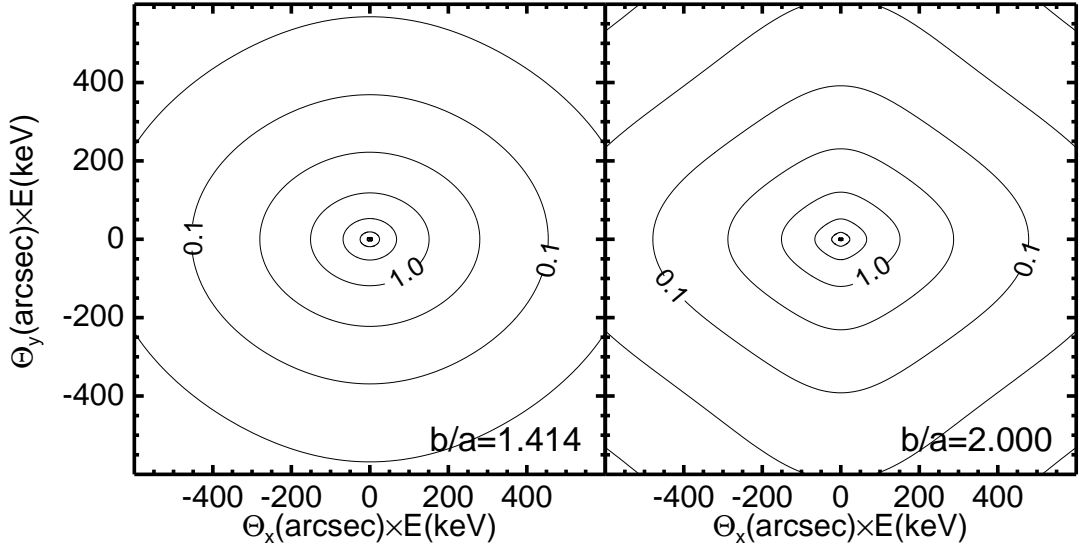}
\caption{\label{fig:contours for uniformly-distributed dust mix}
        \footnotesize
	%f15.eps = fFFF15.eps:
        Contours of constant $I(\theta,\phi)/I(\theta_N,0)=10^{-n/2}$,
	for $\theta_N\equiv150\arcsec$,
	for uniformly-distributed
	mixture of partially-aligned grains if silicate grains
	have axial ratio (a) $b/a=\sqrt{2}$; (b) $b/a=2$.
	Contours are labelled by $I/I(\theta_N,0)$.
	Contours plotted are for $E=1\keV$; for other energies,
	the halo intensities can be estimated from the approximate
	scaling relation $\tilde{\sigma}(E,\Theta)\approx
	\tilde{\sigma}(1\keV,\Theta E/\keV)$
	}
\end{center}
\end{figure}

\begin{figure}[ht]
\begin{center}
\includegraphics[width=16.0cm,angle=0]{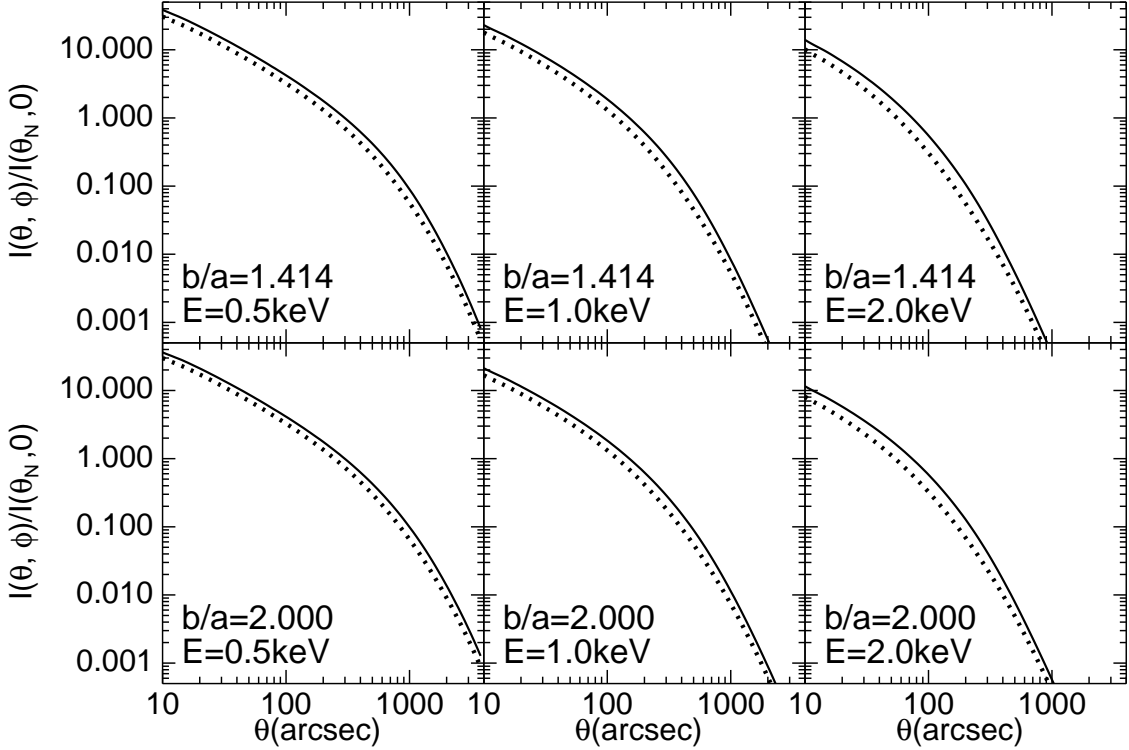}
\caption{\label{fig:I(theta,phi) at selected phi for uniform dust}
         \footnotesize
	$I(\theta,\phi)/I(\theta_N,0)$ vs.\ $\theta$ for
	$\phi=0$ (solid curve) and $\phi=\pi/2$ (dotted curve)
	for uniformly-distributed grains
	at $E=0.5,1,2\keV$, for assumed silicate axial ratio
	$b/a=\sqrt{2}$ and $b/a=2$.
	The arbitrary normalization point
	$\theta_N\equiv 150{\rm\,arcsec}/E(\keV)$.
	}
\end{center}
\end{figure}

\begin{figure}[ht]
\begin{center}
\hspace*{-1.5em}
\includegraphics[width=8.7cm,angle=0]{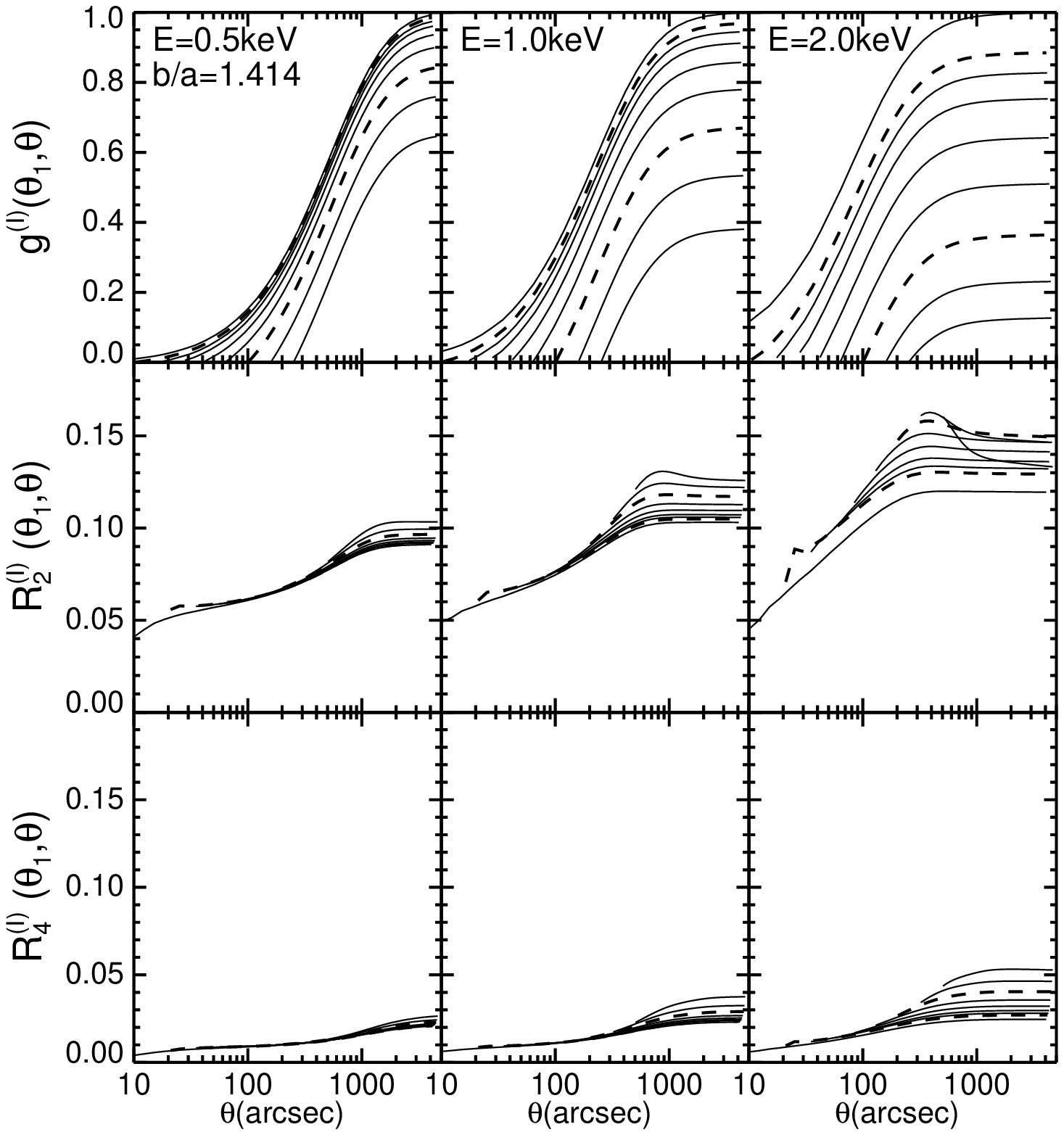}
\hspace*{-2em}\includegraphics[width=8.7cm,angle=0]{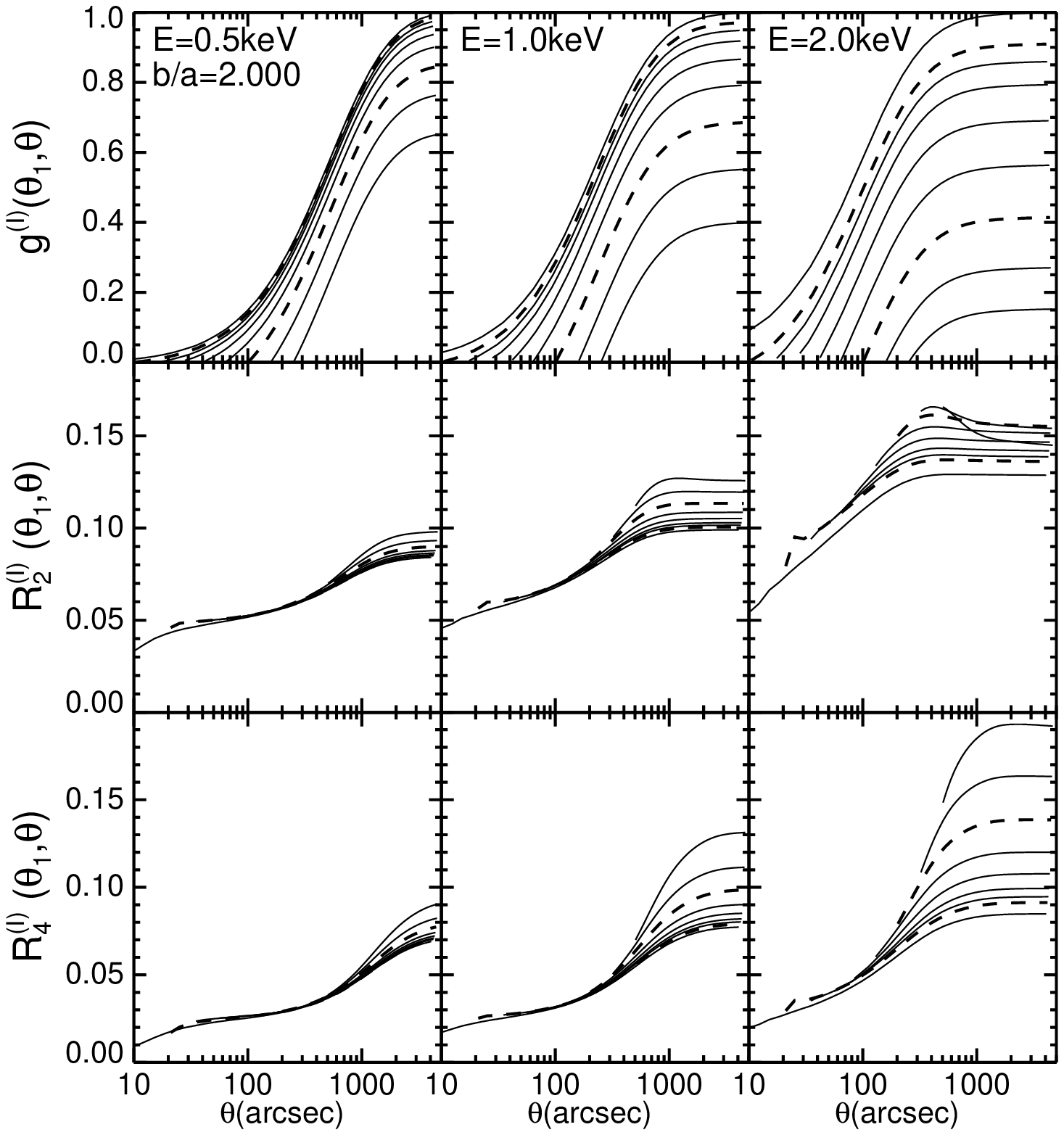}
\caption{\label{fig:R_2,R_4: E=0.5,1,2keV; mix; uniform; b/a=1.4,2}
        \footnotesize
	%f17a.eps = fE17a.eps, f17b.eps = fE17b.eps .
        $g^{(I)}(\theta_1,\theta)$, the fraction of the scattered energy
	in the annulus $[\theta_1,\theta]$,
	and the asymmetry parameters $R_2^{(I)}(\theta_1,\theta)$ 
	and $R_4^{(I)}(\theta_1,\theta)$ vs.\ $\theta$
	for the annulus $[\theta_1,\theta]$, 
        for grains uniformly-distributed between observer and
        source, for $E=0.5,1,2\keV$, and where the silicate
	grains have been assumed to have axial ratio
	$b/a=\sqrt{2}$ and 2.
	}
\end{center}b
\end{figure}

Fig.\ \ref{fig:dsigma/dOmega vs. theta for size dists.} shows the
differential scattering cross section per H nucleon as a function of
scattering angle $\theta$, for $\phi=0$ and $\phi=90^\circ$,
assuming the magnetic field to be in the $\bxhat$
direction ($\phi=0$).  
For $\theta>0$, the scattering is stronger for $\phi=0$ (the direction of
the short axis of the aligned grains), but the differences between
$\phi=0$ and $\phi=90^\circ$ are of course reduced compared to the
single-grain results in Fig.\ \ref{fig:dsigmadomega perp to los}
because we are averaging over an extended size distribution, and except for
the largest grains, there is only partial grain alignment.
Nevertheless, the differences can be as large as $\sim$40\% 
%\btdnote{Khosrow: is this statement numerically correct?  (It is difficult
%for me to read this off the figures.)}
at
$\theta\gtsim 200(\keV/E)~{\rm arcsec}$.

Fig.\
\ref{fig:R_2,R_4: 0.5,1,2keV; mix; sheet; b/a=1.4,2}
shows the anisotropy measures $R_2^{(\sigma)}(0,\theta)$ and 
$R_4^{(\sigma)}(0,\theta)$ 
for the partially-aligned grain models with $b/a=\sqrt{2}$ and 2.
Note that, for fixed $\theta E$, the anisotropy measures $R_2$ and
$R_4$ are larger at higher energies.
The higher $Z$ elements in the 
silicate grains (versus $Z=6$ for the carbon grains) cause $|m-1|$ to decline
less rapidly with energy than for carbon grains; as a result,
the silicate grains provide an increasing fraction of the total scattering
as the energy is increased.  Since the silicate grains are aligned, but the
carbon grains are not, the anisotropy measures $R_2$ and $R_4$ therefore
increase with increasing $E$.

The top panel of 
Fig.\ 
\ref{fig:R_2,R_4: 0.5,1,2keV; mix; sheet; b/a=1.4,2}
shows
$g(0,\Theta)$, the fraction of the scattered power having scattering angles
$<\Theta$, calculated for $E=0.5,1,2\keV$ and the grain models for
$b/a=\sqrt{2}$ and 2. 
If the dust is located at a single distance $D_d$ from the
observer, not too close to the source, the anisotropy measures of the
observed halo are
directly related to the anisotropy measures $R_\ell^{(\sigma)}$ of the
dust mixture:
\beq
R_\ell^{(I)}(\theta_1,\theta_2) = 
R_\ell^{(\sigma)}\left(\frac{\theta_1}{\zeta},\frac{\theta_2}{\zeta}\right)
\eeq
where $\zeta=(D-D_d)/D$, where $D_d$ is the distance of the dust from the
observer. 
Thus far we have been considering scattering by dust at a single distance.
This will apply, for example, if most of the dust on the line-of-sight
to the source is concentrated in a single cloud and the source-cloud
distance was large compared to the
extent of the cloud along the line-of-sight.
It also applies to scattering
by Galactic dust of photons from an AGN, QSO, or GRB, in which case 
$\zeta\rightarrow1$.

\subsection{Uniformly-Distributed Dust}

If the dust is distributed between the observer and the source, the
scattering halo depends on the specific dust distribution.
As a simple example, we consider dust with uniform density for
$0.01 \leq z/D \leq 1$, with zero density for $z < .01D$.\footnote{
    The dust-free zone near the source is introduced in order to
    keep $I(0,0)$ finite.  For dust uniformly-distributed all the
    way to the source, it can be seen that $I(\theta,\phi)\propto \theta^{-1}$
    for $\theta\rightarrow 0$.  This divergence is of little practical
    consequence, because the enclosed power 
    $\int_0^\theta I 2\pi\theta^\prime d\theta^\prime\rightarrow 0$ as
    $\theta\rightarrow 0$.
    }

Fig.\ 
\ref{fig:contours for uniformly-distributed dust mix}
shows contours of scattered intensity for this case.
In contrast to Fig.\ \ref{fig:halo images}, the isointensity contours are
no longer perfect ellipses for this case,
as the result of the scattering contributions from the carbon spheres and
nonaligned silicate spheroids.
Nevertheless, it is seen that the isointensity contours are noticeably
noncircular.
Fig.\ 
\ref{fig:I(theta,phi) at selected phi for uniform dust}
shows the normalized scattered intensity in the $\phi=0$ and
$\phi=90^\circ$ directions for the two partially-aligned mixtures.

The anisotropy measures $R_2^{(I)}(\theta_1,\theta_2)$ and 
$R_4^{(I)}(\theta_1,\theta_2)$ for uniformly-distributed partially-aligned
dust are shown in Fig.\ 
\ref{fig:R_2,R_4: E=0.5,1,2keV; mix; uniform; b/a=1.4,2},
for the two values of $b/a$.
Results are shown for a number of choices of $\theta_1$, the inner
radius of the annulus over which $I(\theta,\phi)$ is assumed to be
measured.
The function $g(\theta_1,\theta_2)$, shown in the top panel, gives
the fraction of the total scattered power that falls in the annulus
$[\theta_1,\theta_2]$.  The observer will want to ensure that the
chosen annulus $[\theta_1,\theta_2]$ will have enough scattered signal in
it to allow reliable determination of the moment $R_2$ and, ideally,
$R_4$.

\section{\label{sec:discussion}
         Discussion}
\subsection{Observational Considerations}

We will assume that there is no uncertainty regarding the position of
the point source on the sky; 
this position is taken to be the origin of a polar coordinate system
$(\theta,\phi_\obs)$ 
with the direction of $\phi_\obs=0$ taken to be whatever
coordinate system is convenient to the observer.
We further assume that the image $I(\theta,\phi_\obs)$
has had the instrumental point spread function
(p.s.f.) subtracted; this is especially important if the p.s.f.\ is not
azimuthally symmetric.
Real images may contain extraneous background or foreground sources
that need to be recognized and removed; we assume that standard
methods are used to interpolate the scattered intensity 
$I(\theta,\phi_\obs)$ in such regions.  In the following discussion
it is assumed that $I(\theta,\phi_\obs)$ is the observed
scattered intensity.

An annulus $[\theta_1, \theta_2]$ 
is chosen that contains
a strong scattered signal, with $\theta_1$ chosen to avoid the central
core that will be dominated by the instrumental p.s.f., and $\theta_2$
chosen to ideally extend beyond the radius where the scattered halo
intensity begins to significantly decline.
If $I(\theta,\phi_\obs)$ is the observed intensity (corrected for nonuniform
backgrounds), a new coordinate system $\phi\equiv\phi_\obs-\Delta$ is
defined, with the rotation angle $\Delta$ determined by the condition
\beqa \label{eq:condition for phi=0}
\Delta &=& \frac{1}{2}\arctan(A/B)
~~~,
\\
A &\equiv& \int_{\theta_1}^{\theta_2} d\theta\sin\theta
           \int_0^{2\pi}d\phi_\obs 
           I(\theta,\phi_\obs)\sin(2\phi_\obs)
~~~,
\\
B &\equiv& \int_{\theta_1}^{\theta_2} d\theta\sin\theta
           \int_0^{2\pi}d\phi_\obs 
           I(\theta,\phi_\obs)\cos(2\phi_\obs)
~~~.
\\
\eeqa
In this coordinate system $(\theta,\phi)$, the anisotropy statistics
$R_\ell^{(I)}(\theta_1,\theta_2)$ can be calculated from 
eq.\ (\ref{eq:R_ell^(I)}).
The resulting $R_2^{(I)}(\theta_1,\theta_2)$ and $R_4^{(I)}(\theta_1,\theta_2)$
can be compared to the theoretical predictions in Figure
\ref{fig:R_2,R_4: 0.5,1,2keV; mix; sheet; b/a=1.4,2}
(for dust at a single distance) or
Figure \ref{fig:R_2,R_4: E=0.5,1,2keV; mix; uniform; b/a=1.4,2}
(for uniformly-distributed dust).

We have calculated the anisotropy statistics $R_\ell^{(I)}$ for
plane-parallel distributions of dust:
dust in a sheet at a single distance and dust uniformly distributed between 
observer and source.
In real situations, gradients in the dust density distribution perpendicular
to the line-of-sight (on linear scales of order 
$\theta_{\rm halo}\approx 1\arcmin\times1\kpc\approx0.3\pc$)
can contribute to the $R_\ell^{(I)}$.  
How can halo anisotropies
due to aligned dust grains be separated from anisotropies due to dust density
gradients?
\begin{enumerate}
\item The halo anisotropy due to dust grain alignment will have 
      $R_1^{(I)}=R_3^{(I)}=0$.
      Dust density gradients, on the other hand, 
      will generally contribute to all of the
      $R_\ell^{(I)}$, 
      and in fact would be expected to make the largest contribution
      to $R_1^{(I)}$.
      Therefore the magnitude of $R_1^{(I)}$ and $R_3^{(I)}$ 
      can be used to estimate
      the contribution of dust density gradients to $R_2^{(I)}$.
\item The halo anisotropy due to aligned dust grains will be aligned with
      the direction of starlight polarization:
      polarized starlight has ${\bf E}$ parallel to the short axis of the
      aligned grains.  Therefore the scattered halo will have its major
      axis parallel to the direction of starlight polarization:
      the angle $\Delta$ should
      coincide with the direction of starlight polarization. 
\item If the anisotropic signal $R_2^{(I)}$ is from aligned grains,
      the angle $\Delta$ will be insensitive to the choice of
      annulus $[\theta_1,\theta_2]$.
      If $\Delta$ is found to vary significantly from one annulus to
      another, the observed anisotropy will have an appreciable
      component from some source other than aligned dust grains.
      -- nonuniform dust, or other X-ray sources
      in the field.
\end{enumerate}

Ideally, the X-ray scattering halo would be observed for a source where
the optical starlight polarization is known either for the source itself
or a nearby stellar companion.  In some cases, the X-ray source may not
be bright enough at optical wavelengths to permit polarization
measurements.  However, the polarization of starlight
is known to display large-scale coherence over the sky
(see, e.g., Mathewson \& Ford 1970); therefore if starlight polarization
has been observed for a star that is nearby on the sky, with a similar
amount of reddening, it is reasonable to presume that the dust in front
of the candidate X-ray source would produce a similar degree of
optical polarization.

\begin{figure}[h]
\includegraphics[angle=270,width=16cm]{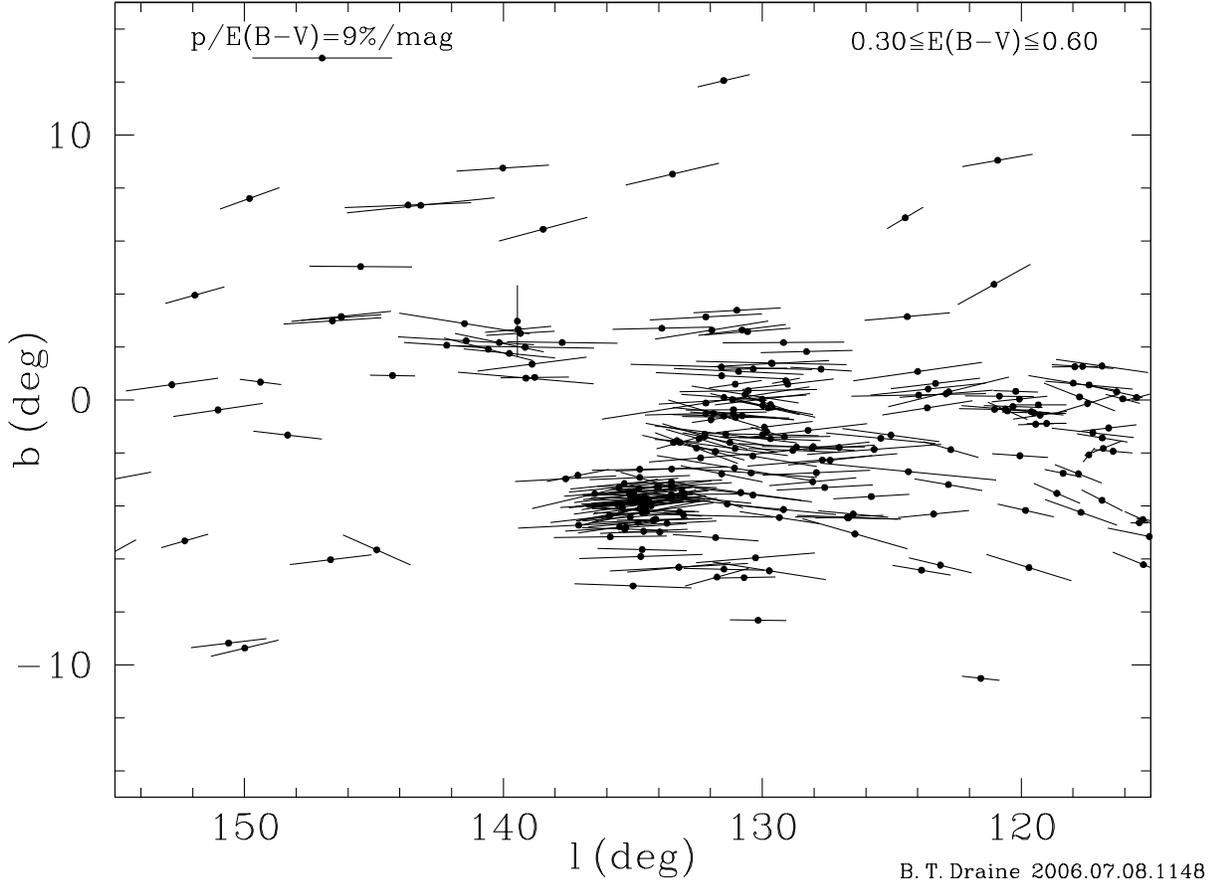}
\caption{\footnotesize
         \label{fig:polmap}
         Line segments show ratio direction of
	 starlight polarization, with length proportional to
	 $p(V)/E(B-V)$, where
	 $p(V)$ is the polarization, and $E(B-V)$ is the reddening.
	 Data from Heiles (2000).
	 In this region of the sky, the dust grain alignment is
	 evidently quite uniform, and the polarization per unit
	 $E(B-V)$ is $\sim2/3$ of the empirical maximum.
	 }
\end{figure}

Directions with large observed starlight polarization tend to lie
close to the galactic plane (so that significant amounts of reddening will
be present) and in directions perpendicular to the general direction of
the local magnetic field.
An example of a region 
with large observed starlight polarization per unit $E(B-V)$ is
$l = 130 \pm 10\deg$.  Figure \ref{fig:polmap} shows $p(V)/E(B-V)$ for
stars in a section of the Galactic plane near $l\approx 130\deg$.
The stars shown are limited to $0.30 \leq E(B-V) \leq 0.60$~mag
(i.e., $0.9 \ltsim A_V \ltsim 1.9$~mag).
The large-scale coherence of the polarization direction
is apparent, and it is also evident that
the ratio of polarization to reddening is relatively uniform.  This indicates
that the magnetic field is relatively well-ordered, and that the efficiency
of grain alignment is also relatively uniform.  An X-ray source in this
general region would presumably be observed through the same aligned dust
as the stars for which the polarization has already been measured.

\subsection{Predictions}

We have considered two models for interstellar grains, both with
partially aligned silicate oblate spheroids, but for two different axial
ratios.  The aligned grains that produce polarization of starlight
also produce anisotropic X-ray scattering.  
While we have explicitly calculated the anisotropic X-ray scattering
only for the optimal case where the magnetic field $\bB$ is unidirectional
and perpendicular to the line-of-sight, we expect that for other
magnetic field geometries, the anisotropy $R_2^{(I)}$ will decrease in
proportion to $\pmax/E(B-V)$, because for optically-thin conditions
both $R_2^{(I)}$ and $\pmax/E(B-V)$ respond similarly to changes in the
magnetic field direction.\footnote{
   E.g., both are unchanged if $\bB\rightarrow-\bB$; both vanish if
   $\bB\parallel\bzhat$; and both vanish if 50\% of the dust has
   symmetry axis
   $\bahat_1\parallel\bxhat$, and 50\% has $\bahat_1\parallel\byhat$.
   }
We can therefore predict the
ratio of the fractional anisotropy of the X-ray halo 
to the ratio of the optical polarization to total reddening.

We suppose that it is possible to measure the X-ray halo for
$\theta_1<\theta<\theta_2$, with $\theta_1\ltsim 200\arcsec$, and
$\theta_2 \gtsim 300\arcsec$.
If the dust is distributed approximately uniformly between observer
and source,
from Figure \ref{fig:R_2,R_4: E=0.5,1,2keV; mix; uniform; b/a=1.4,2}
we see that for
X-ray energy $E = 1\keV$, 
\beq
\label{eq:result for R2/(pmax/E(B-V))}
R_2^{(I)}(\theta_1,\theta_2)
\gtsim 0.09 \times
\left[
\frac{\pmax}{0.09E(B-V)/{\rm mag}}
\right]
~~~,
\eeq
provided the annulus has $\theta_1\ltsim 300\arcsec$,
and $\theta_2\gtsim 300\arcsec$.
The two cases $b/a=\sqrt{2}$ and 2 give similar values for 
the quadrupole anisotropy $R_2^{(I)}$ -- increased grain oblateness
is offset by reduced degree of partial alignment.
As discussed above, the anisotropy $R_2^{(I)}$ is an increasing function
of energy over the range 0.5--2~keV because the relative contribution of
the nonspherical silicate grains to the scattering increases with energy.

It is not critical that the dust be distributed uniformly for the result
(\ref{eq:result for R2/(pmax/E(B-V))}) to be applicable, only that a
negligible amount of the starlight polarization be produced by dust
that is very close to the source, as the anisotropic 
X-ray halo produced by this
dust may be largely lost at small angles 
unless it is possible to use small values of $\theta_1$.

Aligned grains are also predicted to produce an octupole
anisotropy in the X-ray halo,
measured by $R_4^{(I)}$.  Unfortunately, this anisotropy
can be washed out by moderate rotation of the magnetic field direction along
the line-of-sight.
However, if the starlight polarization has 
$\pmax\approx 0.09 E(B-V)/{\rm mag}$,
then we can suppose that the magnetic field in the dusty regions is 
approximately perpendicular to the line-of-sight and unidirectional
(otherwise 
the net starlight polarization would be reduced).  In this case, we can
use the statistic $R_4^{(I)}$ as a test of grain models.  The model with
minimal axial ratio $b/a\approx1.4$ for the spheroids predicts
$R_4^{(I)}\approx 0.03$ at 1~keV (see Figure 
\ref{fig:R_2,R_4: E=0.5,1,2keV; mix; uniform; b/a=1.4,2}).
If larger values of $R_4^{(I)}$ are observed, it will suggest more extreme
grain shapes (for example, 
$R_4^{(I)}\approx 0.10$ would be consistent with $b/a\approx 2$
for the silicate spheroids, but inconsistent with $b/a\approx\sqrt{2}$).
Therefore, measurement of $R_4^{(I)}$ has the potential to
discriminate between grain models with different grain shape.

\section{\label{sec:summary}
         Summary}
The principal results of this paper are as follows:
\begin{enumerate}
\item We show that anomalous diffraction theory can be used for accurate
      calculation of X-ray scattering from dust grains 
      with arbitrary shape.  Fast Fourier transforms may be employed for
      efficient calculation of the scattering halo.
\item Differential scattering cross sections are calculated for oblate 
      spheroids, and are shown to have substantial anisotropies if the
      spheroid axial ratio differs appreciably from unity.
\item Models of spherical carbonaceous grains and partially-aligned
      silicate spheroids are found that reproduce the observed interstellar
      extinction and polarization of starlight as a function of wavelength.
      Confirming a result found previously by Kim \& Martin (1995), we find
      that for silicate axial ratio $b/a\approx\sqrt{2}$ 
      the model can reproduce
      the largest observed starlight polarization, 
      provided that the silicate spheroids
      with $\aeff\gtsim0.1\micron$ are almost perfectly aligned (with their
      short axis parallel to the magnetic field).
      For larger axial ratios, the grain alignment need not be so complete.
\item X-ray scattering halos are calculated for aligned interstellar dust
      grains.  For realistic size distributions 
      and fractional
      alignments,
      the scattered halo shows substantial anisotropy.
\item We propose statistics $R_\ell^{(I)}(\theta_1,\theta_2)$ to measure
      the halo anistropy in an annulus $\theta_1<\theta<\theta_2$.
      We predict the values of $R_2^{(I)}/[\pmax/E(B-V)]$.
      We find that $R_2^{(I)}$ is large enough to be
      measured on sightlines to X-ray point sources 
      where the grains are aligned so as to
      produce starlight polarization with $\pmax/E(B-V)\gtsim 0.05/{\rm mag}$.
\item We show that the octupole anisotropy $R_4^{(I)}$ is sensitive to the
      assumed grain shape.  On favorable sightlines, $R_4^{(I)}$ can be used
      to constrain the geometry of interstellar grains.
\end{enumerate}
\acknowledgments
This research was supported in part by NSF grants AST-0216105 and
AST-0406883.
We are grateful to an anonymous referee for helpful comments.
B.T.D. is grateful to R. Bandiera for valuable discussions, 
and to R.H. Lupton for availability of the SM
software package.

\appendix
\section{Least-squares fitting procedure}
Let $n_c(a)$ be the number density of grains with composition $c$ and
radii $\leq a$.  It is convenient to define $u\equiv\ln a$ and
$y_c$ such that $dn_c/d\ln a=dn_c/du \equiv \exp(y_c)$.
Then $dn_c/da$ is positive definite for $-\infty < y_c < \infty$.

The size distribution is sampled at $N_{\rm rad}$ sizes,
and extinction and polarization are calculated at $N_\lambda$
wavelengths.

The terms $\Psi_j$ include the following contributions
\beqa
\Psi_1 &=& \alpha_1 
         \left[
               \frac{V_{\rm sil}}{V_{\rm sil,0}} -1
         \right]
\\
\Psi_2 &=& \alpha_2 
         \left[
               \frac{V_{\rm car}-V_{\rm PAH}}{V_{\rm car,0}} -1
         \right]
\\
\Psi_{2+j} &=& \frac{\alpha_3}{N_\lambda^{1/2}} 
         \left[
               \frac{A_{\rm mod}(\lambda_j)}{A_\obs(\lambda_j)} -1 
         \right]
         ~~~j=1,...,N_\lambda
\\
\Psi_{N_\lambda+2+j} &=& \frac{\alpha_4}{N_\lambda^{1/2}}
         \left[
         \frac{p_{\rm mod}(\lambda_j)}{p_\obs(\lambda_j)} -1
         \right]
         ~~~j=1,...,N_\lambda
\\
\Psi_{2N_\lambda+2+j} &=& \frac{\alpha_5}{(N_{\rm rad}-1)^{1/2}}
         \left[\min\left( \left(\frac{d\ln f}{du}\right)_{j+1/2} , 
	   0 \right)\right]^2
         ~~~j=1,...,N_{\rm rad}-1
\\
\Psi_{2N_\lambda+N_{\rm rad}+2}   &=& \alpha_6
          \left[\max\left( f(a_{N_{\rm rad}})  - 1, 0\right)\right]^2
\\
\Psi_{2N_\lambda+N_{\rm rad}+1+j} &=& \frac{\alpha_7}{(N_{\rm rad}-2)^{1/2}}
         \left[ \frac{d^2y_{\rm sil}}{du^2} \right]
	  ~~~a=a_j
          ~~~j=2,...,N_{\rm rad}-1
\\
\Psi_{2N_\lambda+2N_{\rm rad}-1+j} &=& \frac{\alpha_8}{(N_{\rm rad}-2)^{1/2}}
         \left[ \frac{d^2y_{\rm car}}{du^2} \right]
	  ~~~a=a_j
          ~~~j=2,...,N_{\rm rad}-1
\\
\Psi_{2N_\lambda+3N_{\rm rad}-3+j} &=& \frac{\alpha_9}{(N_{\rm rad}-2)^{1/2}}
         \left[ \frac{d^2\ln f}{du^2} \right]
	  ~~~a=a_j
          ~~~j=2,...,N_{\rm rad}-1
\eeqa

The radii $a_j$ are assumed to be uniformly distributed in $u=\ln a$:
$u_{j+1}-u_j=\Delta u$.
The derivatives are evaluated using the usual differencing:
$(df/du)_{j+1/2}=(f_{j+1}-f_j)/\Delta u$ and
$(d^2f/du^2)_j=(f_{j+1}+f_{j-1}-2f_j)/(\Delta u)^2$.

Here $V_{\rm sil,0}=2.29\times10^{-27}\cm^3/{\rm H}$ 
and $V_{\rm car,0}=1.57\times10^{-27}\cm^3/{\rm H}$ 
are target values based on the
estimated elemental abundances of species such as Mg, Si, Fe, and C,
the fractions of these elements believed to be in solid form, and
assumed densities for the silicate and carbonaceous material.
For the target volume for silicate grains 
we assume a composition
Mg$_{1.1}$Fe$_{0.9}$SiO$_4$, with Mg/H=$3.3\times10^{-5}$,
and density $\rho=3.7\g\cm^{-3}$.  
This
would consume 87\%, 93\%, 94\%, and 29\% of the 
current estimated solar composition values
for Mg, Fe, Si, and O, respectively
[(Mg/H)$_\odot=3.8\times10^{-5}$ (Grevesse \& Sauval 1998), 
(Fe/H)$_\odot=2.9\times10^{-5}$ (Asplund et al.\ 2000),
(Si/H)$_\odot=3.2\times10^{-5}$ (Asplund 2000),
and
(O/H)$_\odot=4.6\times10^{-4}$ (Asplund et al.\ 2004)].
For the target volume of carbonaceous grains we assume
70\% of the current estimated solar abundance
(C/H)$_\odot=2.46\times10^{-4}$ (Asplund et al.\ 2005) to be
in grains with $\rho=2.2\g\cm^{-3}$.

The coefficients $\alpha_1$ and $\alpha_2$ weight the penalties for
deviating from the target abundances;
$\alpha_3$ and $\alpha_4$ weight the penalties for fractional errors
in reproducing the observed extinction and polarization;
$\alpha_5$ weights the penalty if the alignment fraction $f(a)$ is not
a monotonically-increasing function of size $a$;
$\alpha_6$ and $\alpha_7$ penalize non-smoothness in the size distributions;
and 
$\alpha_8$ penalizes non-smoothness in the alignment function.

%\begin{table}
%\begin{center}
%\caption{Penalty Coefficients}
%\begin{tabular}{c c}
%\hline
%  $\alpha_1$ & \\
%  $\alpha_2$ & \\
%  $\alpha_3$ & \\
%  $\alpha_4$ & \\
%  $\alpha_5$ & \\
%  $\alpha_6$ & \\
%  $\alpha_7$ & \\
%\hline
%\end{tabular}
%\end{center}
%\end{table}
%--------------------------------------------------------------------------

\begin{thebibliography}{}
\bibitem[Asplund 2000]{As00}
        Asplund, M.
	2000,
	A\&A, 359, 755
\bibitem[Asplund etal 2000]{ANT00}
        Asplund, M., Nordlund, \AA, Trampedach, R.,
	Allende Prieto, C., \& Stein, R.F. 2000,
	A\&A, 359, 729
\bibitem[Asplund+Grevesse+Sauval_etal_2004]{AGS04}
        Asplund, M., Grevesse, N., Sauval, A.J., Allende Prieto, C.,
	\& Kiselman, D.
	2004,
	A\&A, 417, 751
\bibitem[Asplund+Grevess+Sauval_etal_2005]{AGS05}
        Asplund, M., Grevesse, N., Sauval, A.J., Allende Prieto, C.,
	\& Blomme, R.
	2005, 
	A\&A, 431, 693
\bibitem[Bohlin et al 1978]{BSD78}
        Bohlin, R.C., Savage, B.D., \& Drake, J.F. 1978,
	ApJ, 224, 132

\bibitem[Bohren & Huffman 1983]{BH83}
        Bohren, C.F., \& Huffman, D.R. 1983,
	Absorption and Scattering of Light by Small Particles
	(NY: Wiley).

\bibitem[Catura 1983]{Ca83}
        % Evidence for X-ray scattering by interstellar dust
        Catura, R.C. 1983,
        ApJ, 275, 645

\bibitem[Costantini et al 2005]{CFP05}
        % Absorption and scattering by interstellar dust: XMM-Newton
        % observations of Cyg X-2
        Costantini, E., Freyberg, M.J., \& Predehl, P.,
	A\&A, 444, 187

%\bibitem[Cunha et al 2006]{CHL06}
%        % Ne Abundances in B-sars of the Orion Association: Solving
%        % the Solar Model Problem?
%        Cunha, K., Hubeny, I., \& Lanz, T. 2006,
%	ApJ, in press (astro-ph/0606738).

\bibitem[Debye 1909]{De09}
        Debye, P. 1909,
	Ann. Phys., 30, 57

\bibitem[Draine 2003a]{Dr03a}
        % Interstellar Dust Grains
        Draine, B.T. 2003a,
	ARAA, 41, 241

\bibitem[Draine 2003b]{Dr03b}
	% Scattering by Interstellar Dust Grains. II. X-Rays
	Draine, B.T. 2003b,
	ApJ, 598, 1026

\bibitem[Draine & Li 1984]{DL84}
        % Optical properties of interstellar graphite and silicate grains
        Draine, B.T., \& Lee, H.-M. 1984,
	ApJ, 285, 89

\bibitem[Draine \& Malhotra 1993]{DM93}
        % On Graphite and the 2175 A Extinction Profile
        Draine, B.T., \& Malhotra, S. 1993,
	ApJ, 414, 632

\bibitem[Draine \& Tan 2003]{DT03}
        % The X-ray scattering halo around nova cygni 1992: testing a
        % model for interstellar dust
        Draine, B.T., \& Tan, J.C. 2003,
	ApJ, 594, 347

\bibitem[Draine \& Weingartner 1997]{DW97}
        % Radiative Torques on Interstellar Grains. II. Grain Alignment
        Draine, B.T., \& Weingartner, J.C. 1997,
	ApJ, 480, 633

\bibitem[Frisch+Dorschner+Geiss_etal_1999]{FGL99}
        Frisch, P.C., Dorschner, J.M., Geiss, J.,
	Greenberg, J.M., Gr\"un, E., Landgraf, M., Hoppe, P., et al.,
	1999,
	ApJ, 525, 492

\bibitem[Grevesse+Sauval_1998]{GS98}
        Grevesse, N., \& Sauval, A.J.,
	1998,
	Sp.\ Sci.\ Rev., 85, 161

\bibitem[Hall 1949]{Ha49}
        Hall, J.S. 1949,
	Science, 109, 166

\bibitem[Hayakawa 1970]{Ha70}
        Hayakawa, S. 1970,
        Prog. Theor. Phys., 43, 1224

\bibitem[Heiles 2000]{He00}
        % 9286 Stars: An Agglomeration of Stellar Polarization Catalogs
        Heiles, C. 2000,
	AJ, 119, 923

\bibitem[Hiltner 1949]{Hi49}
        Hiltner, W.A. 1949,
	Science, 109, 165

\bibitem[Kim & Martin 1995]{KM95}
        % The size distributionof interstellar dust particles
        % as determined from polarization: spheroids
        Kim, S.-H., \& Martin, P.G. 1995,
	ApJ, 442, 172

\bibitem[Lazarian & Draine 1999a]{LD99a}
        % Thermal Flipping and Thermal Trapping -- New Elements in Grain
        % Dynamics
        Lazarian, A., \& Draine, B.T. 1999a,
	ApJ, 516, L37

\bibitem[Lazarian & Draine 1999b]{LD99b}
        % Nuclear Spin Relaxation Within Interstellar Grains
        Lazarian, A., \& Draine, B.T. 1999b,
	ApJ, 520, L67

\bibitem[Li & Draine 2001]{LD01}
        Li, A., \& Draine, B.T. 2001,
	ApJ 554, 778

\bibitem[Martin et al 1992]{MAW92}
        %Interstellar polarization from 3 to 5 microns in reddened stars
        Martin, P.G., Adamson, A.J., Whittet, D.C.B.,
	Hough, J.H., Bailey, J.A., Kim, S.-H.,
	Sato, S., Tamura, M., \& Yamashita, T. 1992,
	ApJ, 392, 691

\bibitem[Mathewson & Ford 1970]{MF70}
        % Polarization observations of 1800 stars
        Mathewson, D.S., \& Ford, V.L. 1970,
	Mem. R. Astron. Soc., 74, 139

\bibitem[Mathis et al 1995]{MCFK95}
        Mathis, J.S., Cohen, D., Finley, J.P., \& Krautter, J. 1995,
	ApJ, 449, 320

\bibitem[Mathis et al 1977]{MRN77}
        Mathis, J.S., Rumpl, W., \& Nordsieck, K.H. 1977,
	ApJ, 217, 425

\bibitem[Mauche \& Gorenstein 1986]{MG86}
        % Measurements of X-ray scattering from interstellar grains
        Mauche, C.W., \& Gorenstein, P. 1986,
	ApJ, 302, 371

\bibitem[Mie 1908]{Mi08}
        Mie, G. 1908,
	Ann. Phys. 25, 377

\bibitem[Mishchenko 2000]{Mi00}
        % Calculation of the amplitude matrix for a nonspherical
        % particle in a fixed orientation
        Mishchenko, M.I. 2000,
	Applied Optics, 39, 1026

\bibitem[Mishchenko \& Travis 1994]{MT94}
        % T-matrix computations of light scattering by large spheroidal
        % particles
        Mishchenko, M.I., \& Travis, L.D. 1994,
        Opt. Comm., 109, 16

\bibitem[Mishchenko et al 1996]{MTM96}
        % T-matrix computations of light scattering by nonspherical
        % particles:  a review
        Mishchenko, M.I., Travis, L.D., \& Mackowski, D.W. 1996,
	JQSRT, 55, 535

\bibitem[Overbeck 1965]{Ov65}
        % Small-angle scattering of celestial X-rays by interstellar grains
        Overbeck, J.W. 1965,
        ApJ, 141, 864

\bibitem[Predehl \& Schmitt 1995]{PS95}
        % X-raying the interstellar medium: ROSAT observations of dust
        % scattering hallos
        Predehl, P., \& Schmitt, J.H.M.M. 1995,
	A\&A, 293, 889

\bibitem[Purcell 1979]{Pu79}
        % Suprathermal rotation of interstellar grains
        Purcell, E.M. 1979,
	ApJ, 231, 404

\bibitem[Serkowski 1973]{Ser73}
        Serkowski, K. 1973,
	in Interstellar Dust and Related Topics.
	IAU Symposium 52.
	Ed. J.M. Greenberg and H.C. van de Hulst.
	(Dordrecht: Reidel)
	p. 145

\bibitem[Serkowski et al 1975]{SMF75}
        % Wavelength dependence of interstellar polarization and
        % ratio of total to selective extinction
        Serkowski, K., Mathewson, D.L., \& Ford, V.L. 1975,
	ApJ, 196, 261

\bibitem[Slysh 1969]{Sl69}
        Slysh, V.I. 1969,
        Nature, 224, 159

\bibitem[Smith, Edgar, \& Shafer 2002]{SES02}
        % The X-ray halo of GX 13+1
        Smith, R.K., Edgar, R.J., \& Shafer, R.A. 2002,
	ApJ, 581, 562

\bibitem[Sofia \& Meyer 2001]{SM01}
        % Interstellar Abundance Standards Revisited
        Sofia, U.J., \& Meyer, D.M. 2001,
	ApJ, 554, L221

\bibitem[Temperton 1983]{Te83}
        % Self-sorting mixed radix Fast Fourier Transforms
        Temperton, C.J. 1983,
        J. Comp. Phys, 52, 1

\bibitem[Temperton 1992]{Te92}
        % A Generalized Prime Factor FFT Algorithm for Any N=2^p3^q5^r
        Temperton, C. 1992,
        J. Scientific and Statistical Computing, 13, 676

\bibitem[van de Hulst 1957]{vdH57}
        van de Hulst, H.C. 1957,
	Light Scattering by Small Particles,
	(New York: Wiley)

\bibitem[Vanlandingham etal 2005]{VSS05}
        Vanlandingham, K.M., Schwartz, G.J., Shore, S.N.
	Starrfield, S., \& Wagner, R.M. 2005,
	ApJ, 624, 914
	
\bibitem[Waterman 1971]{Wa71}
        Waterman, P.C. 1971,
	Phys. Rev. D 3, 825

\bibitem[Weingartner & Draine 2001]{WD01}
        % Dust grain size distributions and extinction in the Milky Way,
        % LMC, and SMC
        Weingartner, J.C., \& Draine, B.T. 2001,
	ApJ, 548, 296 (WD01)

\bibitem[Whittet et al 1992]{WMH92}
        Whittet, D.C.B., Martin, P.G., Hough, J.H., Rouse, M.F.,
	Bailey, J.A., \& Axon, D.J. 1992,
	ApJ, 386, 562

\bibitem[Wielaard et al 1997]{WMM97}
        % Improved T matrix computations for large nonabsorbing and
        % weakly absorbing nonspherical particles and comparison with
        % geometryc optics approximation
        Wielaard, D.J., Mishchenko, M.I., Marke, A., \& Carlson, B.E. 1997,
	Appl. Opt., 36, 4305

\bibitem[Wiscombe (1980)]{Wi80}
	Wiscombe, W.J. 1980, 
	Appl. Opt., 19, 1505

\bibitem[Wiscombe (1996)]{Wi96}
	Wiscombe, W.J. 1996, 
	NCAR Technical Note NCAR/TN-140+STR,
	ftp://climate.gsfc.nasa.gov/pub/wiscombe/Single-Scatt/Homogen\_Sphere/Exact\_Mie/NCARMieReport.pdf

\bibitem[Woo et al 1994]{WCD94}
        Woo, J.W., Clark, G.C., Day, C.S.R., Nagase, F., \& Takeshima, T. 1994,
	ApJ, 436, L5

\bibitem[Yao etal 2003]{YZZF03}
        Yao, Y., Zhang, S.N., Zhang, X.L., \& Feng, Y.X. 2003,
	ApJ, 594, L43

\bibitem[Zubko et al 2004]{ZDA04}
        Zubko, V., Dwek, E., \& Arendt, R.G. 2004,
	ApJS, 152, 211

\end{thebibliography}
\end{document}